\documentclass[twocolumn, twocolappendix]{aastex701}

\newcommand{\diracuw}{Dept. of Astronomy \& the DiRAC Institute, University of Washington, Box 351580, Seattle, WA 98195, USA}
\newcommand{\nau}{Department of Astronomy and Planetary Science, Northern Arizona University, Flagstaff, USA}

\usepackage{comment}

\shorttitle{Hyrax: Rapid ML Experimentation \& Unsupervised Discovery for Astronomy}
\shortauthors{Ghosh, Oldag, Tauraso, et al.}

\font\bngxi=bang10 scaled 1100

\def\*#1*#2{o\null{#2}{#1}}

\def\d#1{\oalign{\smash{#1}\crcr\hidewidth{$\!$\rm.}\hidewidth}}

\def\sh#1{\setbox0=\hbox{#1}%
     \kern-.02em\copy0\kern-\wd0
     \kern.04em\copy0\kern-\wd0
     \kern-.02em\raise.0433em\box0 }

\begin{document}

\title{Hyrax: An Extensible Framework for Rapid ML Experimentation and Unsupervised Discovery in the Era of Rubin, Roman, and Euclid}

\correspondingauthor{A. Ghosh: LSST-DA Catalyst Fellow}

\author[orcid=0000-0002-2525-9647, gname=Aritra, sname=Ghosh]{Aritra Ghosh ({\bngxi Airt/r \*gh*eaSh})}
\altaffiliation{A. Ghosh, D. Oldag \& M. Tauraso contributed equally to this work and are listed alphabetically as co-first-authors.}
\affiliation{\diracuw}
\affil{eScience Institute, University of Washington, Seattle, WA, U.S.A.}
\email{aritrag@uw.edu; aritraghsh09@gmail.com}

\author[orcid=0000-0001-6984-8411, gname=Drew, sname=Oldag]{Drew Oldag}
\altaffiliation{A. Ghosh, D. Oldag \& M. Tauraso contributed equally to this work and are listed alphabetically as co-first-authors.}
\affiliation{\diracuw}
\email{awoldag@uw.edu}

\author[orcid=0009-0008-6215-1783, gname=Michael, sname=Tauraso]{Michael Tauraso}
\altaffiliation{A. Ghosh, D. Oldag \& M. Tauraso contributed equally to this work and are listed alphabetically as co-first-authors.}
\affiliation{\diracuw}
\email{mtauraso@uw.edu}


\author[0000-0001-5576-8189]{Andrew J. Connolly}
\affiliation{\diracuw}
\affil{eScience Institute, University of Washington, Seattle, WA, U.S.A.}
\email{ajc@astro.washington.edu}

\author[0000-0001-6957-1627]{Peter Ferguson}
\affiliation{\diracuw}
\affil{eScience Institute, University of Washington, Seattle, WA, U.S.A.}
\email{pferguso@uw.edu}

\author[0009-0006-2411-723X]{Derek Jones}
\affiliation{\diracuw}
\email{dtj@mac.com}

\author[0000-0002-3475-7648]{Gourav Khullar}
\altaffiliation{Baum Postdoctoral Fellow for Innovative Astronomy}
\affiliation{\diracuw}
\affil{eScience Institute, University of Washington, Seattle, WA, U.S.A.}
\email{gkhullar@uw.edu}

\author[0000-0001-7357-0889]{Argyro Sasli}
\affiliation{School of Physics and Astronomy, University of Minnesota, Minneapolis, MN 55455, USA}
\email{asasli@umn.edu}

\author[0009-0000-7254-8495]{Samarth Venkatesh}
\affiliation{\diracuw}
\email{samnsid7@uw.edu}

\author[0009-0009-7931-4507]{Gracia Wang}
\affiliation{\diracuw}
\email{wqy37@uw.edu}

\author[0009-0003-3171-3118]{Maxine West}
\affiliation{\diracuw}
\email{maxwest@uw.edu}


\author[0009-0000-5333-9970]{Dylan Berry}
\affiliation{\diracuw}
\email{dberry04@uw.edu}

\author[0000-0003-3287-5250]{Neven Caplar}
\affiliation{\diracuw}
\email{ncaplar@uw.edu}

\author[0000-0001-7335-1715]{Colin Orion Chandler}
\email{coc123@uw.edu}
\affiliation{\diracuw}
\affiliation{\nau}

\author[0009-0007-7376-8853]{Tanawan Chatchadanoraset}
\affiliation{\diracuw}
\email{tchatc24@uw.edu}

\author[0000-0002-8262-2924]{Michael W. Coughlin}
\affiliation{School of Physics and Astronomy, University of Minnesota, Minneapolis, MN 55455, USA}
\email{cough052@umn.edu}

\author[0000-0002-1074-2900]{Melissa DeLucchi}
\affiliation{McWilliams Center for Cosmology and Astrophysics, Department of Physics, Carnegie Mellon University, Pittsburgh, PA 15213, USA}
\email{delucchi@andrew.cmu.edu}

\author[0000-0002-9380-7983]{Alexandra Junell}
\affiliation{School of Physics and Astronomy, University of Minnesota, Minneapolis, MN 55455, USA}
\email{ajunell@umn.edu}

\author[0009-0009-9486-3053]{Diego Miura}
\affil{Department of Astronomy, Yale University, 219 Prospect Street, New Haven, CT 06511, USA}
\email{diego.miura@yale.edu}

\author[0009-0007-3755-0021]{Naomi Morato}
\affiliation{\diracuw}
\email{nmorato@uw.edu}

\author[0000-0001-7129-1325]{Felipe Fontinele Nunes}
\affiliation{School of Physics and Astronomy, University of Minnesota, Minneapolis, MN 55455, USA}
\email{fonti007@umn.edu}

\author[0000-0002-8766-6627]{Hurum Maksora Tohfa}
\affiliation{\diracuw}
\email{htohfa@uw.edu}


\author[0009-0003-1791-8707]{Wilson Beebe}
\affiliation{\diracuw}
\email{wbeebe@uw.edu}

\author[0009-0009-7822-7110]{Doug Branton}
\affil{\diracuw}
\email{brantd@uw.edu}

\author[0009-0007-9870-9032]{Sandro Campos}
\affiliation{McWilliams Center for Cosmology and Astrophysics, Department of Physics, Carnegie Mellon University, Pittsburgh, PA 15213, USA}
\email{scampos@andrew.cmu.edu}

\author[]{Liam Cunningham}
\affil{Department of Physics and Astronomy and PITT PACC, University of Pittsburgh, Pittsburgh, PA 15260, USA}
\email{LDC39@pitt.edu}

\author[0000-0002-5995-9692]{Mi Dai}
\affil{Department of Physics and Astronomy and PITT PACC, University of Pittsburgh, Pittsburgh, PA 15260, USA}
\email{mi.dai@pitt.edu}

\author[0009-0009-2281-7031]{Jeremy Kubica}
\affiliation{McWilliams Center for Cosmology and Astrophysics, Department of Physics, Carnegie Mellon University, Pittsburgh, PA 15213, USA}
\email{jkubica@andrew.cmu.edu}

\author[0000-0001-7179-7406]{Konstantin Malanchev}
\affil{McWilliams Center for Cosmology and Astrophysics, Department of Physics, Carnegie Mellon University, Pittsburgh, PA 15213, USA}
\email{malanchev@cmu.edu}

\author[0000-0003-2271-1527]{Rachel Mandelbaum}
\affil{McWilliams Center for Cosmology and Astrophysics, Department of Physics, Carnegie Mellon University, Pittsburgh, PA 15213, USA}
\email{rmandelb@andrew.cmu.edu}

\author[0009-0005-8764-2608]{Sean McGuire}
\affil{McWilliams Center for Cosmology and Astrophysics, Department of Physics, Carnegie Mellon University, Pittsburgh, PA 15213, USA}
\email{seanmcgu@andrew.cmu.edu}

\author[0000-0002-7075-9931]{Imad Pasha}
\affil{Dragonfly Focused Research Organization, 150 Washington
Avenue, Santa Fe, 87501, NM, USA }
\affil{Department of Astronomy, Yale University, 219 Prospect Street, New Haven, CT 06511, USA}
\email{imad.pasha@yale.edu}

\author[0000-0001-6268-1882]{Dan S. Taranu}
\affil{Department of Astrophysical Sciences, Princeton University, Princeton, NJ 08544, USA}
\email{dtaranu@princeton.edu}

\author[0000-0002-5596-198X]{Tianqing Zhang}
\affil{Department of Physics and Astronomy and PITT PACC, University of Pittsburgh, Pittsburgh, PA 15260, USA}
\email{tq.zhang@pitt.edu}

\begin{abstract}
The NSF-DOE Vera C. Rubin Observatory, Roman Space Telescope, Euclid, and other next-generation surveys will deliver imaging, spectroscopic, and time-domain data at scales that increasingly shift the bottleneck in astronomical machine learning (ML) projects from model design to infrastructure. We present Hyrax, an open-source, modular, GPU-enabled Python framework that supports the full ML lifecycle in astronomy: from data acquisition and training to inference and experiment comparison, with capabilities including multimodal dataset support, integrated vector databases for similarity search, and interactive two- and three-dimensional latent-space exploration for unsupervised discovery. We demonstrate Hyrax's versatility through five representative applications on real survey data: (i) unsupervised representation learning on ${\sim}4\times10^5$ Rubin Legacy Survey of Space and Time (LSST) Data Preview 1 (DP1) galaxies, surfacing new merger and low-surface-brightness candidates missing from reference Euclid and Dark Energy Survey catalogs, while also isolating imaging artifacts---all without labeled training data; (ii) hybrid density-based clustering for identifying cluster-scale gravitational lens candidates in DP1 data; (iii) multimodal early-time transient classification in the Zwicky Transient Facility leveraging light curves, spectra, images, and metadata; (iv) supervised false-positive filtering in shift-and-stack searches for distant solar system objects in the Dark Energy Camera Ecliptic Exploration Project survey; and (v) supervised detection of semi-resolved dwarf galaxies in Hyper Suprime-Cam and LSST-like imaging using synthetic source injection. Together, these results demonstrate that Hyrax provides astronomy-specific ML infrastructure that enables systematic discovery and rapid methodological iteration across next-generation astronomical surveys.
\end{abstract}



\section{Introduction}
\begin{figure}[htbp]
    \centering
    \includegraphics[width=0.8
    \linewidth]{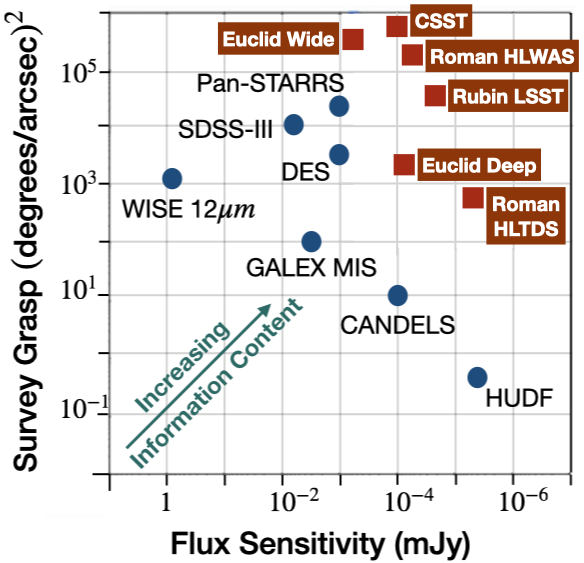}
    \caption{Survey grasp---defined as survey area scaled by angular resolution---is plotted against flux sensitivity for a selection of representative astronomical imaging surveys. Legacy surveys are shown as blue circles, while upcoming wide-fast-deep surveys (e.g., Rubin-LSST, Euclid, Roman, and CSST) are shown as red squares. The diagonal arrow indicates the direction of increasing information content, which scales with both depth and area. This next generation of ground- and space-based surveys will be among the most information-rich to date, enabling data-driven discovery across a broad range of astrophysical domains. Surveys shown are the Cosmic Assembly Near-infrared Deep Extragalactic Legacy Survey \citep[CANDELS;][]{Grogin2011, Koekemoer2011}, Dark Energy Survey \citep[DES;][]{DESCollaboration2016}, Hubble Ultra Deep Field \citep[HUDF;][]{Beckwith2006}, Galaxy Evolution Explorer Medium Imaging Survey \citep[GALEX MIS;][]{Martin2005}, CSST \citep{csst}, Roman High-Latitude Time-Domain Survey \citep[HLTDS;][]{Rose2021}, Roman High-Latitude Wide-Area Survey \citep[HLWAS;][]{Spergel2015Wide-FieldReport}, Rubin LSST \citep{Ivezic2019LSST:Products}, Sloan Digital Sky Survey \citep[SDSS;][]{York2000}, and the Wide-field Infrared Survey Explorer \citep[WISE;][]{Wright2010}.}
    \label{fig:data_growth}
\end{figure}

The forthcoming decade in astronomy will feature an unprecedented array of large-scale surveys spanning imaging and spectroscopy, whose depth, breadth, and cadence promise to revolutionize our understanding of the Universe. Among the wide-field imaging surveys, this new generation includes the NSF-DOE Vera C.\ Rubin Observatory's Legacy Survey of Space \& Time \citep[Rubin-LSST;][]{Ivezic2019LSST:Products}, the Nancy Grace Roman Space Telescope \citep[Roman;][]{Spergel2015Wide-FieldReport, Akeson2019The2020s}, the Euclid mission \citep{Laureijs2011EuclidReport, Racca2016TheDesign}, and the Chinese Space Station Survey Telescope \citep[CSST;][]{csst}, which will map billions of galaxies across cosmic time at multiple wavelengths. In fact, as shown in Figure \ref{fig:data_growth}, their grasp and sensitivity make them the most information-rich surveys to date, promising transformative scientific returns while simultaneously demanding novel approaches capable of handling the scale and complexity of these datasets. 

\begin{figure}[htbp]
    \centering
    \includegraphics[width=0.8
    \linewidth]{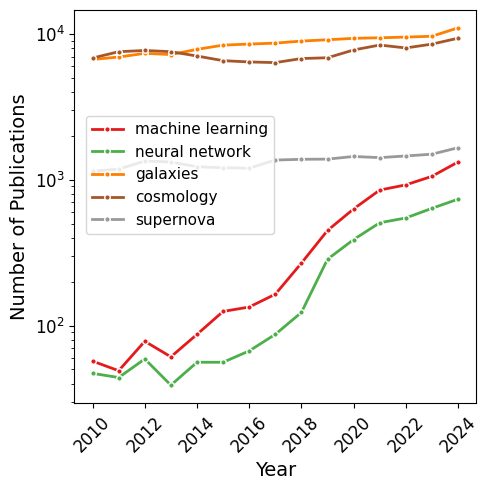}
    \caption{Annual number of peer-reviewed astronomy journal articles indexed in the NASA Astrophysics Data System (ADS) from 2010–2024 containing the indicated keywords in either the title or abstract. The steep, sustained rise in publications referencing ``machine learning" and ``neural network" over the past decade reflects the rapid adoption of these methods across the astronomical community, in contrast to the relatively stable publication rates for evolved sub-domains within astronomy---tracked using the keywords ``galaxies", ``cosmology", and ``supernova". This accelerating interest underscores both the opportunities and the infrastructure challenges that motivated the development of Hyrax.}
    \label{fig:pubs_v_time}
\end{figure}

To prepare for and analyze such datasets, machine learning (ML) has been increasingly employed by astronomers for a wide variety of tasks, with applications spanning galaxy evolution, solar system science, transients, anomaly detection, and exoplanets \citep[e.g.,][; see \cite{Huertas-Company2023TheSurveys} for a review]{dieleman_15, company_15, ml_pz, Shallue2018IdentifyingKepler-90, Duev2019DeepStreaks, gamornet, Walmsley2020GalaxyLearning, Storey-Fisher2020AnomalyNetworks, Villar2021ATransients, Malanchev2021AnomalyDR3, Ghosh2022GaMPEN:Parameters, Muthukrishna2022Real-timeSurveys, Tian2023UsingSurvey, Crenshaw2024UsingSystem, Lochner2025AstronomalyCollaboration}. Reflecting this breadth of application, the number of astronomy publications invoking ML has risen dramatically over the past decade, as shown in Figure \ref{fig:pubs_v_time}. However, most of these efforts have remained project-specific, with each group developing bespoke pipelines that duplicate effort and are difficult to extend or adapt beyond their original scientific context. This fragmentation hampers reuse, slows progress, and makes it challenging to scale methods to the demands of upcoming surveys. Designed to address this gap, \textit{Hyrax is a modular, flexible, and extensible GPU-enabled framework that provides reusable infrastructure for ML projects in astronomy.}

\begin{figure*}[htbp]
    \centering
    \includegraphics[width=0.6\linewidth]{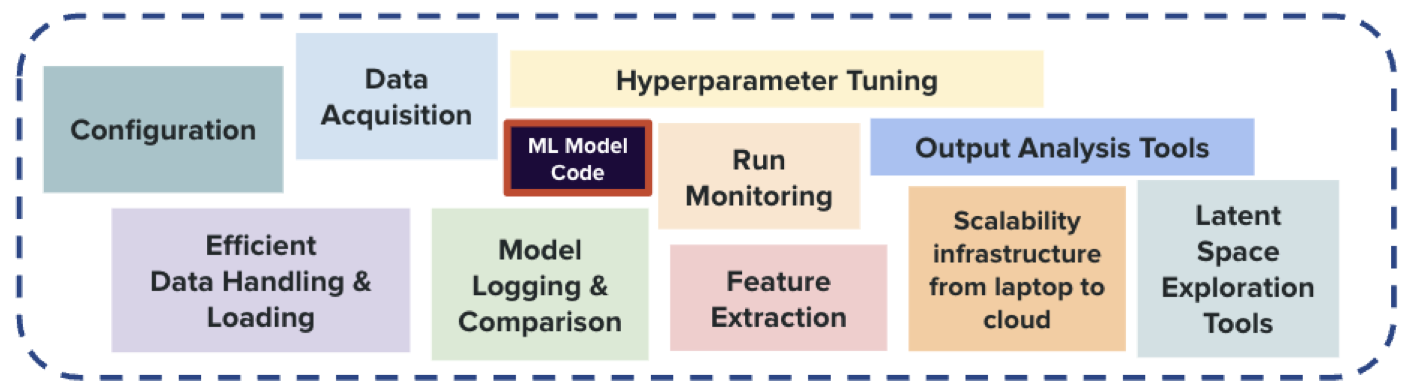}
    \caption{The multitude of different components involved in a typical ML project in astronomy. While the ML model code itself is only one piece of the workflow, successful projects require substantial surrounding infrastructure for tasks such as data acquisition, efficient data loading, run monitoring, and interactive visualization of latent spaces. Hyrax provides robust, astronomy-specific implementations for all of these components---allowing researchers to focus on developing their own models---while also offering a selection of pre-built models for common use cases. This depiction is inspired by the analogous diagram in \citet{Sculley2015HiddenDebt}.}
    \label{fig:design_proposition}
\end{figure*}

ML in astronomy is often framed primarily in terms of model development and evaluation, but in practice, these tasks represent only a small fraction of the overall effort \citep{Sculley2015HiddenDebt, Baylor2017TFX}. The majority of time is typically spent on surrounding tasks: acquiring and organizing data, defining consistent experimental configurations, tracking model outputs, and ensuring that results are reproducible and comparable across iterations. Figure \ref{fig:design_proposition} illustrates this reality: while the ML model itself occupies the central red box, it is surrounded by a wide array of non-model components that are routinely required for a complete ML workflow in astronomy. Astronomers need end-to-end frameworks that support the entire ecosystem of tasks surrounding model development and deployment---including dataset acquisition, preprocessing, efficient data handling, hyperparameter tuning, experiment tracking, scalability, and interactive tools for exploring unsupervised latent spaces. While general-purpose ML libraries such as PyTorch \citep{Ansel2024PyTorchCompilation}, MLflow \citep{mlflow}, and TensorBoard\footnote{\url{https://github.com/tensorflow/tensorboard}} provide powerful building blocks for these components, they are not tailored to the needs of astronomical workflows. Adapting these tools to domain-specific use cases---such as handling Flexible Image Transport System (FITS) images, processing multimodal information for astronomical objects (e.g., spectra, light curves), or integrating with astronomical survey pipelines and data systems---often requires significant time and technical expertise. 

These infrastructure challenges are compounded by practical barriers that limit the accessibility and scalability of ML in astronomy. Scaling ML analysis frameworks from laptop-scale prototypes to high-performance computing systems or cloud computing infrastructure presents newcomers with a steep learning curve as they simultaneously try to assemble a complete ML infrastructure from disparate tools. The recent emergence of astronomical foundation models and associated unsupervised learning approaches \citep[e.g.,][]{Parker2024AstroCLIP:Galaxies, Smith2024AstroPT:Astronomy} introduces additional challenges:  these methods produce high-dimensional latent representations that require specialized tools for similarity search, clustering, and interactive exploration to extract scientific insights. As the surveys shown in Figure \ref{fig:data_growth} come online---with Rubin alone expected to generate over 3 petabytes of images annually \citep[][]{Ivezic2019LSST:Products}---these scalability and usability challenges will intensify. Without accessible, astronomy-tailored infrastructure, the potential for data-driven discovery in these unprecedented datasets risks being limited by technical barriers rather than scientific imagination.

Hyrax addresses these challenges by letting users focus on defining the logic of their ML models, while handling the rest of the workflow through standardized, astronomy-aware infrastructure.  Hyrax includes astronomy-specific tools for each block shown in Figure \ref{fig:design_proposition} and is intentionally application-agnostic, supporting both supervised and unsupervised learning workflows. It includes built-in support for scaling from laptops to multi-GPU high-performance computing (HPC) environments. Hyrax also provides dedicated infrastructure for unsupervised discovery, including interactive latent-space visualization and rapid similarity search, which are largely absent from both general-purpose ML libraries and existing astronomy-focused tools. Hyrax is designed to be modular and extensible, allowing researchers to use as much or as little of the framework as needed for their specific scientific applications.

The remainder of this paper is organized as follows. In \S \ref{sec:hyrax_description}, we outline the core infrastructure of Hyrax, including its modular design, scalable training, and astronomy-specific tools. In \S \ref{sec:applications}, we present previews of five distinct applications that demonstrate Hyrax's versatility across different astronomical domains, spanning from extragalactic astronomy to solar system science. These examples are not intended to present final scientific results, but rather to illustrate how Hyrax's modular framework can be applied to diverse research topics spanning different data types, analysis techniques, and scientific goals. Full scientific analysis for each of these previews will appear in separate follow-up publications. Finally, in \S \ref{sec:summary}, we summarize this work and outline future directions for Hyrax.

\section{What is Hyrax?} \label{sec:hyrax_description}
\begin{figure*}[htbp]
    \centering
    \includegraphics[width=0.9\linewidth]{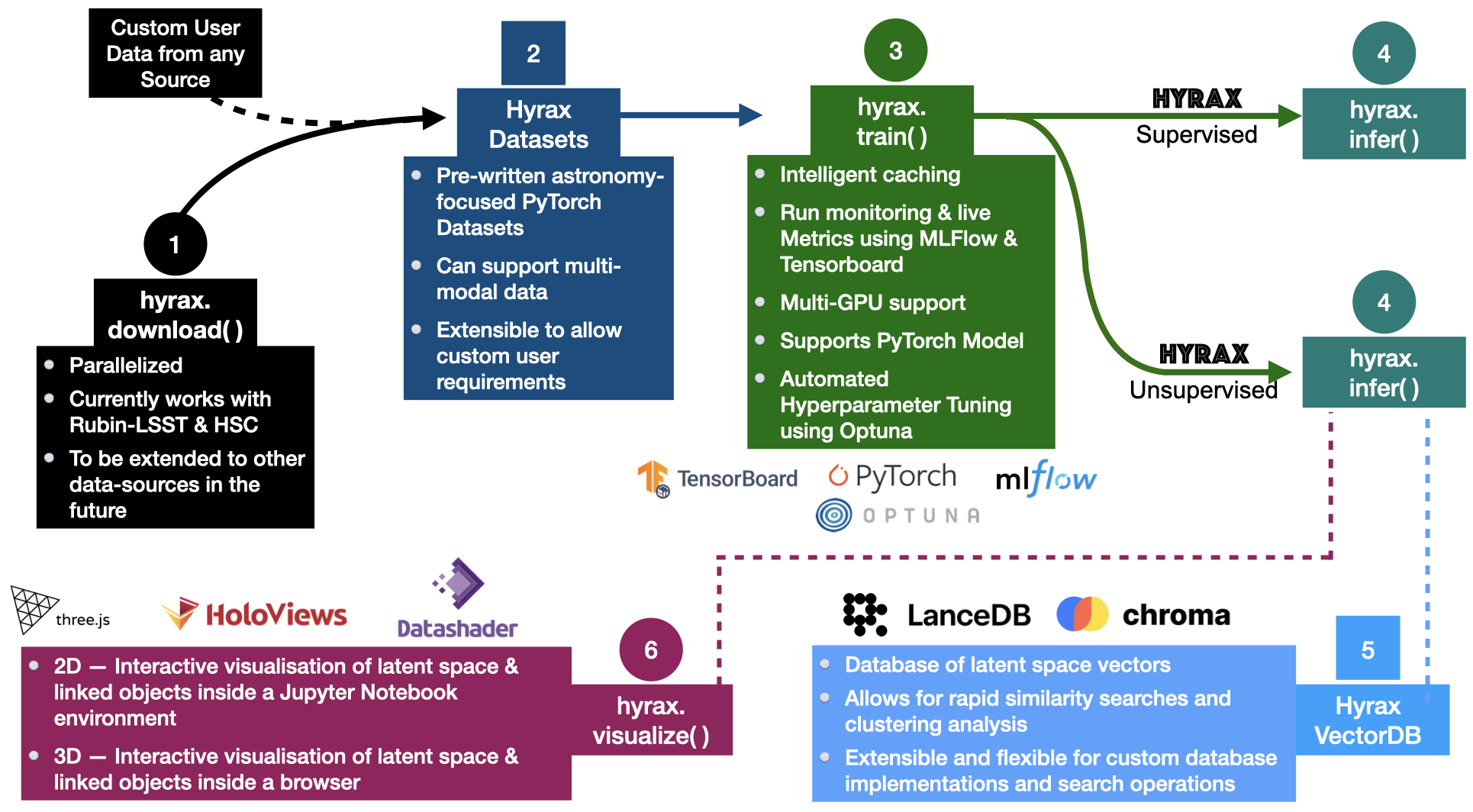}
    \caption{Schematic of the primary modules within Hyrax.                                                                                                         
      (1) The workflow begins with \texttt{hyrax.download()} for parallelized data retrieval from supported surveys or custom user-provided sources.                  
      (2) Retrieved data are organized into flexible, astronomy-tailored ML datasets.                                                                                 
      (3) The training module (\texttt{hyrax.train()}) offers intelligent pre-caching, live monitoring, logging, and multi-GPU support, via e.g., PyTorch, MLflow, Optuna \citep{Akiba2019Optuna}, and TensorBoard,                                
      (4) Both supervised and unsupervised models are supported, with inference for either performed via \texttt{hyrax.infer()}.                                      
      (5) For unsupervised workflows, outputs can be stored in flexible vector databases (e.g., LanceDB\footnote{\url{https://github.com/lancedb/lancedb}},           
  ChromaDB\footnote{\url{https://github.com/chroma-core/chroma}}) to enable rapid similarity search and clustering.                                                     
      (6) Built-in visualization modules provide interactive 2D/3D latent-space exploration (via e.g., three.js\footnote{\url{https://threejs.org/}},                 
  HoloViews\footnote{\url{https://holoviews.org/}}, Datashader\footnote{\url{https://datashader.org/}}), linking model outputs back to source data and facilitating   
  rapid discovery in large survey datasets.}
      \altcaption{A numbered workflow diagram showing the six primary modules of the Hyrax framework connected by directional arrows. Box 1 (left): hyrax.download(), 
      which performs parallelized data retrieval currently supporting Rubin-LSST and HSC, with a dashed arrow indicating that custom user data can also be ingested directly. Box 2: Hyrax Datasets, which organizes retrieved data into pre-written astronomy-focused ML dataset classes that support multimodal data and are extensible to custom user requirements. Box 3 (center): hyrax.train(), which provides intelligent caching, run monitoring and live metrics via MLflow and TensorBoard, multi-GPU support, a PyTorch model interface, and automated hyperparameter tuning via Optuna; logos for all four tools are shown. Box 4 (right):
      hyrax.infer(), split into two parallel paths — one for supervised models and one for unsupervised models — both terminating at hyrax.infer(). Box 5 (bottom right): Hyrax VectorDB, a database of latent-space vectors enabling rapid similarity searches, clustering analysis, and extensible custom database backends; LanceDB and ChromaDB logos are shown. Box 6 (bottom center): hyrax.visualize(), providing 2D interactive latent-space visualization linked to objects within a Jupyter Notebook environment, and 3D interactive visualization in a web browser; three.js, HoloViews, and Datashader logos are shown. Arrows indicate the sequential flow from download through datasets, training, inference, vector database storage, and visualization.}
    \label{fig:hyrax_schematic}
\end{figure*}

Hyrax supports the full lifecycle of ML and unsupervised discovery projects in astronomy. This section describes the core infrastructure components that make this possible. We begin with an overview of Hyrax's design philosophy (\S \ref{sec:hyrax_overview}), then cover data acquisition and organization (\S \ref{sec:data_acquisition}), training and experiment comparison (\S \ref{sec:model_training}), vector database integration for rapid similarity searches (\S \ref{sec:vdb}), and interactive tools for exploring learned latent-space representations (\S \ref{sec:viz}). 
Hyrax has been developed as an open-source framework since its inception and can be installed via \texttt{pip install hyrax}. Links to the source code, documentation, and tutorials are provided in Appendix \ref{sec:ap:code}.

\subsection{Hyrax Overview} \label{sec:hyrax_overview}

Hyrax is built around a small set of high-level workflow ``verbs" (e.g., \texttt{hyrax.train()}, \texttt{hyrax.visualize()}) that correspond to the major stages of an ML analysis. These actions can be executed independently or combined into end-to-end workflows or pipelines. These operations provide users with a consistent organizational structure for defining, executing, and revisiting ML experiments.

Figure \ref{fig:hyrax_schematic} illustrates the primary modules within Hyrax and their interconnections. Hyrax workflows typically begin with parallelized data retrieval from supported astronomical surveys; as of this writing, these include Rubin-LSST and the Hyper Suprime-Cam Subaru Strategic Program \citep[HSC-SSP;][]{Aihara2018TheDesign, Aihara:2022}. Users working with other datasets can bypass this module entirely and directly provide their own data to Hyrax. Retrieved/provided data are organized into flexible, astronomy-tailored ML datasets that can handle multimodal astronomical data (e.g., images, spectra, light curves). The training verb, \texttt{hyrax.train()}, orchestrates model training with intelligent pre-caching to minimize bottlenecks, automatic multi-GPU support, and hyperparameter tuning; it also provides live monitoring of the training through MLflow and TensorBoard. Both supervised and unsupervised models are supported through a unified interface, with inference performed via \texttt{hyrax.infer()}. For unsupervised workflows, Hyrax additionally provides specialized infrastructure, including integrated vector databases for rapid similarity search and interactive visualization tools in two and three dimensions for exploring high-dimensional latent spaces that link back to source data. 

Hyrax builds on well-established open-source tools wherever possible (e.g., PyTorch, MLflow, ChromaDB), and connects them through a layer of astronomy-specific functionality. This includes native support for FITS images, handling of multimodal astronomical data, and direct integration with survey data systems such as the Rubin Observatory Data Butler \citep[][]{Jenness2022TheSystem}. Where existing tools do not adequately address the needs of astronomical workflows, Hyrax introduces custom components, such as software for performing reverse-image/spectra searches. This hybrid approach leverages mature software ecosystems while enabling the reuse of infrastructure across diverse survey-scale astronomical analyses. Beyond supporting custom model development, Hyrax also offers a library of pre-built models for common astronomical use cases, enabling newcomers to quickly build upon established methods while experienced users retain full flexibility to implement their own architectures.

At the same time, Hyrax is designed to be easily extensible. Users can integrate their own models, dataset classes, and visualizations while relying on Hyrax’s shared infrastructure. Over time, we envision a community-driven ecosystem where researchers share and reuse these components, enabling rapid experimentation and collaboration on ML projects across astronomy. 

\subsection{Data Acquisition and Definition} \label{sec:data_acquisition}
\begin{figure*}[htbp]
    \centering
    \includegraphics[width=\linewidth]{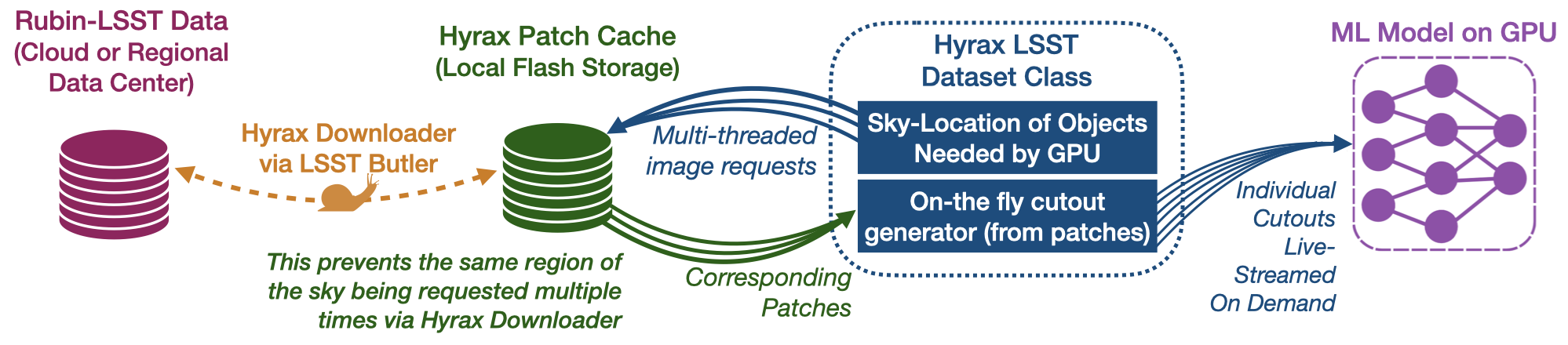}
    \caption{Training ML models on survey-scale imaging data requires efficient access to potentially millions of image cutouts. This block diagram shows how Hyrax's \texttt{LSSTDataset} class addresses this: survey patches are retrieved via the Rubin Butler and cached at the patch level on local storage, preventing repeated queries for the same sky region. During training, image cutouts are \textit{preemptively} generated on the fly from the cached patches via multi-threaded requests and streamed directly to the ML model on the GPU, enabling efficient training in both on-platform (e.g., Rubin Science Platform) and offline environments (e.g., HPC cluster with access to the Rubin Butler).}
    \label{fig:lsst_downloaded_dataset}
\end{figure*}

\begin{figure*}[htbp]
    \centering
    \includegraphics[width=0.95\linewidth]{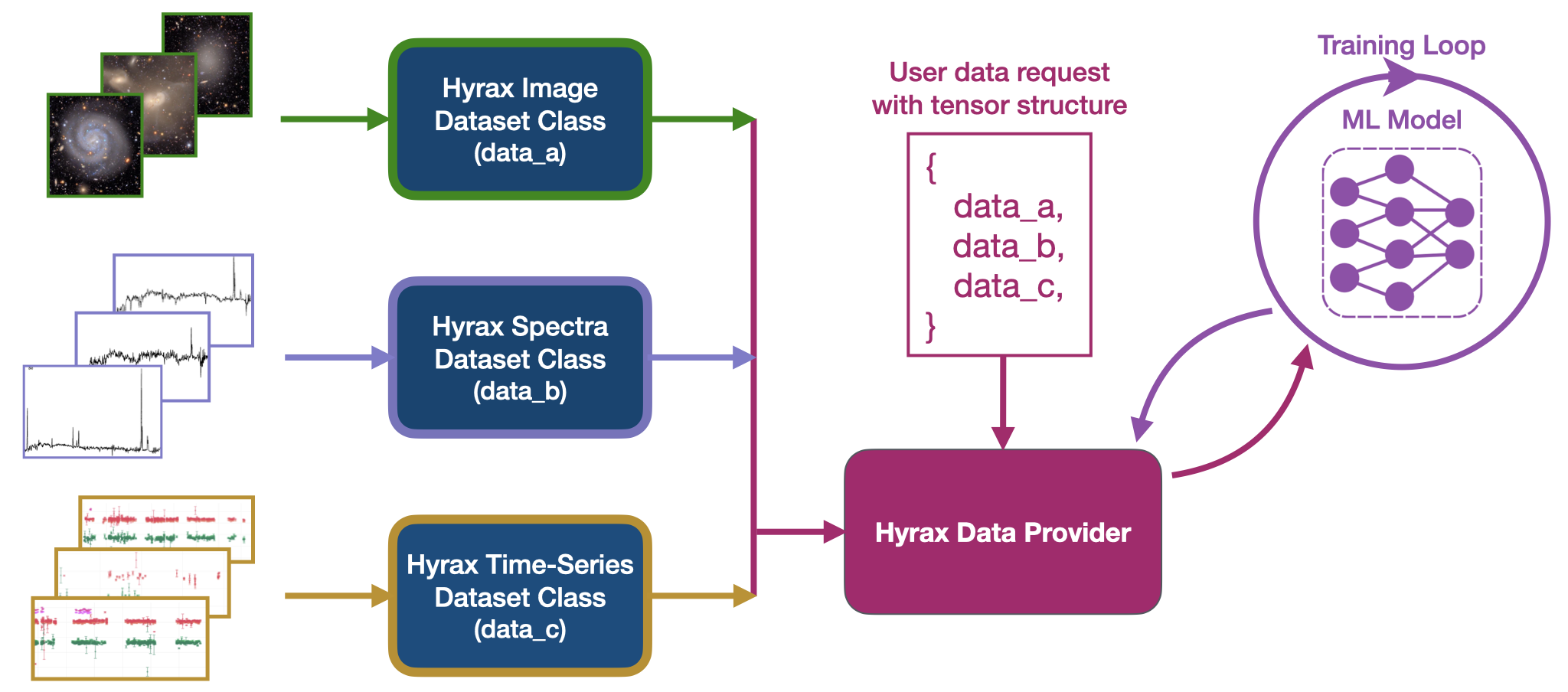}
    \caption{Diagram of the primary components that control the flow of data during training and inference in Hyrax. Data are read by an appropriate \texttt{DatasetClass} that exposes components of the data to \texttt{DataProvider}. The \texttt{DataProvider} prepares a combined dataset based on the user data request configuration. The ML engine will request data samples for training or inference, which are then furnished by \texttt{DataProvider}. The \texttt{DataProvider} can also be configured to preload items from disk and cache them in memory to speed up the training process. In practice, we found that this often sped up training epochs by an order of magnitude.}
    \label{fig:hyraxql}
\end{figure*}

ML workflows in astronomy typically begin with the collection and curation of the data on which models are trained or applied. For much of the past decade, this step has been manageable using relatively simple, survey-specific approaches---for example, downloading data directly from archive web servers using custom scripts or Structured Query Language (SQL) queries. While these methods work well at modest data volumes, they do not scale naturally to the size and complexity of next-generation datasets. Instead, major surveys have introduced dedicated science platforms and data-access ecosystems that enable analysis to be performed close to the data---such as the Rubin Science Platform \citep[RSP;][]{OMullane2021RSP} and the associated Rubin Data Butler, the European Space Agency Datalabs \citep[][]{Navarro2024Datalabs} for Euclid, and the Roman Research Nexus\footnote{\url{https://roman.science.stsci.edu/}}. This shift fundamentally changes how data access and acquisition must be handled in modern ML workflows. 

Hyrax is designed to operate within this evolving data-access landscape by providing configurable tools for efficient data acquisition from large astronomical surveys such as LSST and HSC, with support for additional sources under development. As a concrete example, Figure \ref{fig:lsst_downloaded_dataset} illustrates the LSST-specific dataset class in Hyrax, which manages the flow of data from Rubin servers to the ML models on the GPU while optimizing for rapid data delivery. To do so, Hyrax implements a local caching layer that minimizes repeated requests to the survey servers and enables multi-threaded, on-the-fly generation of cutouts from larger imaging units during training. When local storage resources permit, users may optionally persist these generated cutouts to disk for reuse across experiments. More generally, by streaming only the data required for a given batch directly to the model---rather than requiring bulk download of large imaging datasets---this approach enables efficient ML analyses on survey-scale data without local storage of substantial fractions of the survey. 

In addition to data retrieval, all Hyrax downloaders provide a general set of usability enhancements, including automatic recovery from failed connections, lightweight monitoring to estimate progress and remaining time, and tracking of files that were partially or unsuccessfully acquired. Together, these features reduce friction and the time that astronomers spend on data acquisition and curation, particularly for large samples that require hundreds of thousands or millions of cutouts.

Once data are available, Hyrax treats subsequent stages of the workflow as independent of how those data were acquired: users are \textit{not} required to use the built-in downloaders to take advantage of Hyrax’s training, inference, or analysis infrastructure. Central to this separation is the Hyrax data-definition architecture, shown in Figure \ref{fig:hyraxql}. Dataset classes provide standardized access to individual data modalities---such as images, spectra, or time-series---retrieving specific components from storage given a uniform object identifier. A configurable \texttt{DataProvider} then assembles one or more dataset classes into a structured input based on a user-defined specification, delivering the requested data to ML models on the GPU. This design supports both single-modality inputs and multimodal configurations that combine heterogeneous data products, an increasingly important capability as astronomers have begun to develop foundation and representation-learning models that jointly operate on images, spectra, and time-domain data \citep[e.g.,][]{Rizhko2025AstroMAstronomy,Parker2025AION-1:Sciences,Zhang2024Maven:Science}, alongside dedicated efforts to curate large, survey-scale multimodal datasets explicitly designed to support such approaches \citep[e.g.,][]{TheMultimodalUniverseCollaboration2024TheData}.

\subsection{Model Training, Comparison, and Inference} \label{sec:model_training}

\begin{figure*}[htbp]
    \begin{interactive}{animation}{mlflow_animated.mp4}
    \centering
    \includegraphics[width=0.95\linewidth]{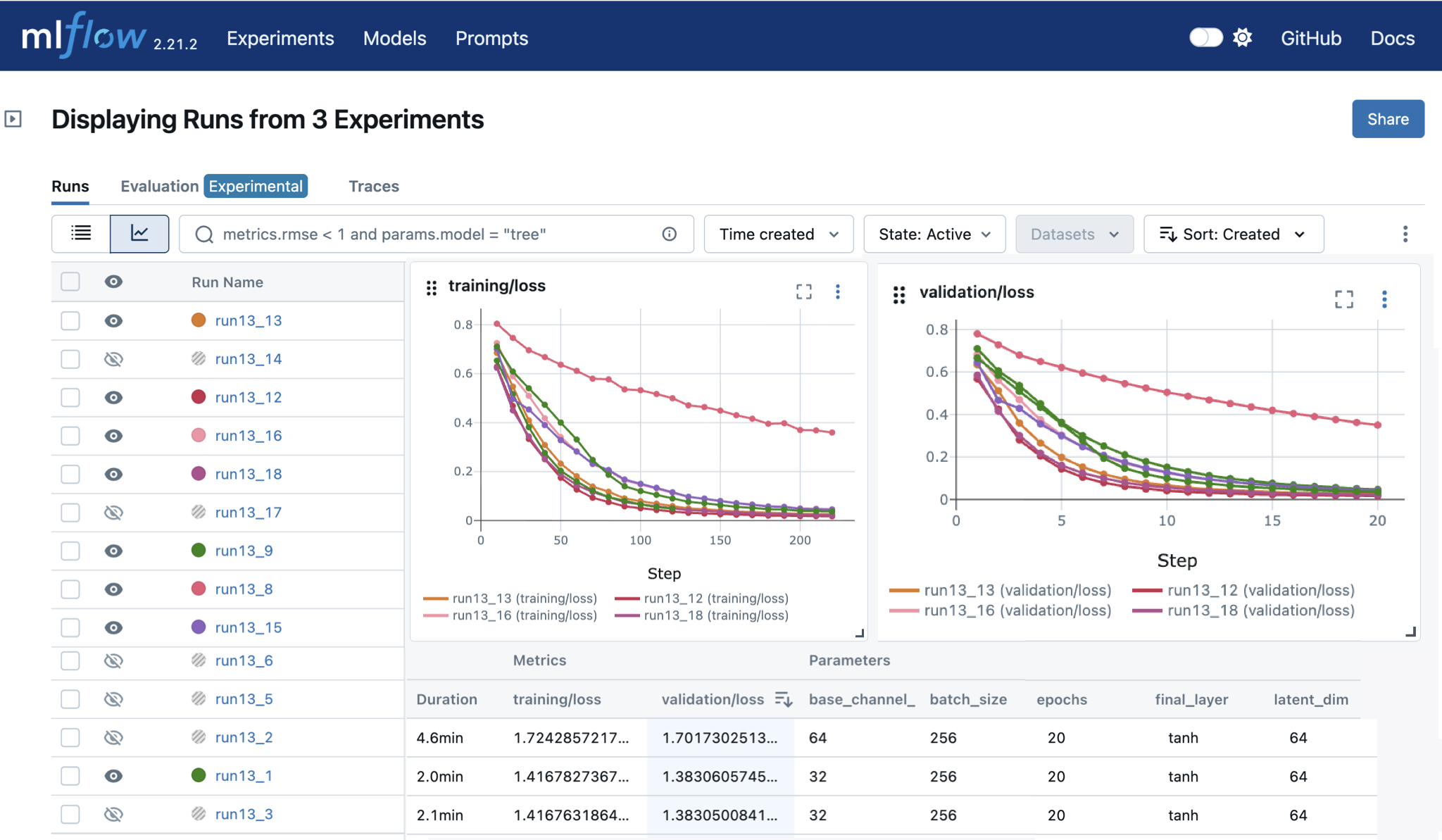}
    \end{interactive}
    \caption{Model comparison and experiment tracking in Hyrax via its native MLflow integration. The static version of this figure shows the MLflow interface populated automatically during Hyrax training runs, displaying loss curves, evaluation metrics, and hyperparameters for multiple experiments executed under a common configuration framework. This enables systematic and easy comparison of models across different datasets and hyperparameter choices without additional user bookkeeping. The animated version of this figure (available in the online version and \href{https://youtu.be/83E7z4bALIQ}{at this link}) demonstrates interactive exploration of this interface, including filtering and sorting runs, inspecting metric histories, and drilling down into individual model configurations during and after training.}
    \label{fig:mlflow}
\end{figure*}

Hyrax centralizes model training and inference within a unified framework that manages data flow, compute resources, and experiment bookkeeping. Users define models, datasets, and training logic through a small set of configuration parameters and code snippets, while Hyrax orchestrates training runs, captures all relevant configuration metadata, and records model outputs in a structured manner on disk. This design allows users to prototype quickly with small datasets locally and then expand to large-scale runs on high-performance computing systems or cloud infrastructure, without requiring changes to user code.

In Hyrax, models and datasets are treated as modular, interchangeable components within the training framework described above. Both are fully customizable: users may integrate their own dataset classes and any PyTorch-based model. Even when defining custom models, users need only specify the core scientific components---such as the model backbone architecture and the logic governing a single training and validation loop---while Hyrax handles training, configuration tracking, and output management. This structured separation enforces consistency across runs, providing the foundation on which systematic exploration and comparison remain practical as the number of experiments grows.

A central component of this iterative workflow is automatic hyperparameter exploration. Hyperparameters---such as learning rates, batch sizes, or architectural choices---strongly influence model behavior, yet are often tuned through informal trial and error. Hyrax enables these choices to be explored through integration with automated hyperparameter optimization frameworks such as Optuna, which systematically explore a predefined parameter space to optimize a chosen evaluation metric. 

As the number of training runs increases, the limiting factor is no longer execution, but comparison. Hyrax addresses this by integrating model comparison directly into the training workflow through native integration with MLflow and TensorBoard. All training runs automatically emit evaluation metrics, hyperparameters, and run metadata to these systems, enabling users to visually and quantitatively compare model performance across experiments without additional bookkeeping.

Figure \ref{fig:mlflow} illustrates this comparison interface in practice for training runs automatically logged by Hyrax via MLflow. Multiple training runs associated with a single experiment can be inspected side-by-side, with direct access to metric histories, configuration details, and run states. In addition to the default training metrics, users can readily record any custom application-specific quantities during training. An accompanying video (see caption) demonstrates the user experience of interactively filtering, sorting, and inspecting runs during and after training. 

Together, automatic hyperparameter exploration and integrated model comparison allow Hyrax users to efficiently probe large parameter spaces and reason about model behavior at scale, transforming large collections of training runs into robust experimental records rather than an unstructured set of outputs.

\subsection{Vector Database and Similarity Search} \label{sec:vdb}
Unsupervised and self-supervised learning methods have become increasingly central to the astronomy ML toolkit as data volumes from new surveys outpace the availability of reliable labels \citep[see][for a review]{Fotopoulou2024AAstronomy}. This shift has been reinforced by the growing use of large foundation and representation-learning models in astronomy---models trained on broad, often unlabeled datasets that produce rich latent embeddings which can be adapted to perform diverse downstream tasks \citep[e.g.,][]{Walmsley2023Zoobot:Morphology, Parker2024AstroCLIP:Galaxies, Smith2024AstroPT:Astronomy, Mohale2024EnablingRepresentations, Zhang2024Maven:Science}. In practice, representations learned in an unsupervised manner tend to organize objects according to meaningful scientific attributes: sources that are nearby in learned latent spaces often share morphological, spectral, or variability characteristics even when these similarities are subtle or challenging to capture with traditional catalog features. Effectively leveraging these representations, therefore, requires tools that enable rapid similarity search and latent-space exploration at scale. Hyrax’s similarity-search toolkit (this section) and our interactive latent-space exploration tools (see \S \ref{sec:viz}) provide precisely these capabilities.

A natural way to explore learned latent representations is through repeated similarity queries, in which a user provides a reference object (e.g., an image, spectrum, or light-curve) and retrieves other sources that lie nearby in the latent space. Such queries support interactive discovery tasks such as identifying analogs or surfacing rare and unusual sources. However, there has been little astronomy-native infrastructure to support this type of exploration at scale. Naïvely computing similarities between a query vector and all other vectors scales as $\mathcal{O}(n)$, while exhaustive pairwise comparisons across the entire dataset scale as $\mathcal{O}(n^2)$, quickly becoming prohibitive for large survey datasets. Hyrax addresses this challenge by storing model outputs in vector databases that index high-dimensional vectors using approximate nearest-neighbor algorithms, enabling efficient similarity search while maintaining high recall.

\begin{figure}[htbp]
    \centering
    \includegraphics[width=0.9\linewidth]{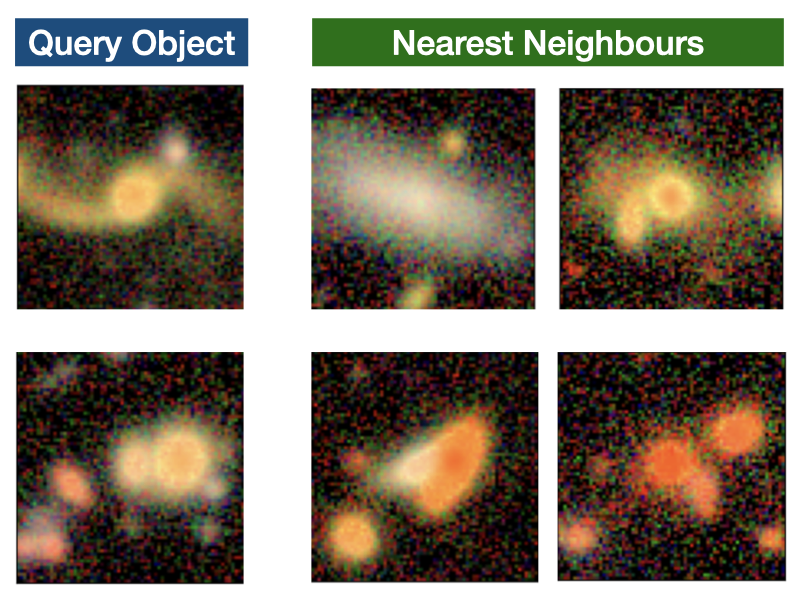}
    \caption{Similarity search examples performed on Rubin DP1 galaxies using an unsupervised ML model within the Hyrax framework (see \S \ref{sec:rubin_dp1_lsst} for details). The left column shows query images, while the right columns show nearest neighbors retrieved via the \texttt{search\_by\_vector()} method. In the top row, a query galaxy exhibiting extended low–surface-brightness features retrieves neighbors with similarly diffuse outer light. In the bottom row, a merger/close-pair system is used as the query, and the retrieved neighbors display comparable interacting/multi-component morphologies. The recovered galaxies share clear, interpretable morphological characteristics with the query sources, demonstrating that similarity searches recover astronomically meaningful structure embedded in the latent space.}
    \label{fig:vdb_demo}
\end{figure}

In Hyrax, the outputs of model inference---such as latent embeddings produced by unsupervised or representation-learning models---are automatically inserted into a user-chosen vector database. Users can then perform similarity queries directly by connecting to the database and using the \texttt{search\_by\_vector()} method, which retrieves the nearest neighbors of a provided query vector from the entire database. Figure \ref{fig:vdb_demo} illustrates this workflow using a latent space learned from Rubin DP1 galaxy images with an unsupervised model (see \S \ref{sec:rubin_dp1_lsst} for details), demonstrating how similarity searches are useful in retrieving astronomically meaningful neighbors in latent space.

At present, Hyrax supports three vector database implementations, LanceDB, ChromaDB, and Qdrant\footnote{\url{https://github.com/qdrant/qdrant}}, via a plugin-based architecture that allows additional backends to be incorporated in the future. All three of these databases rely on the Hierarchical Navigable Small Worlds \citep[HNSW; ][]{Malkov2020HNSW} algorithm to perform searches based on a user-defined distance metric, such as cosine similarity. HNSW-based searches are significantly faster than naive approaches, with a time complexity of $\mathcal{O}(\log n)$, and practical runtimes on the order of milliseconds for databases containing tens of millions of records.

\subsection{Interactive Latent-Space Exploration} \label{sec:viz}
\begin{figure*}[htbp]
    \begin{interactive}{animation}{2d_viz_animated_ffmpeg_compressed.mp4}
    \centering
    \includegraphics[width=0.9\linewidth]{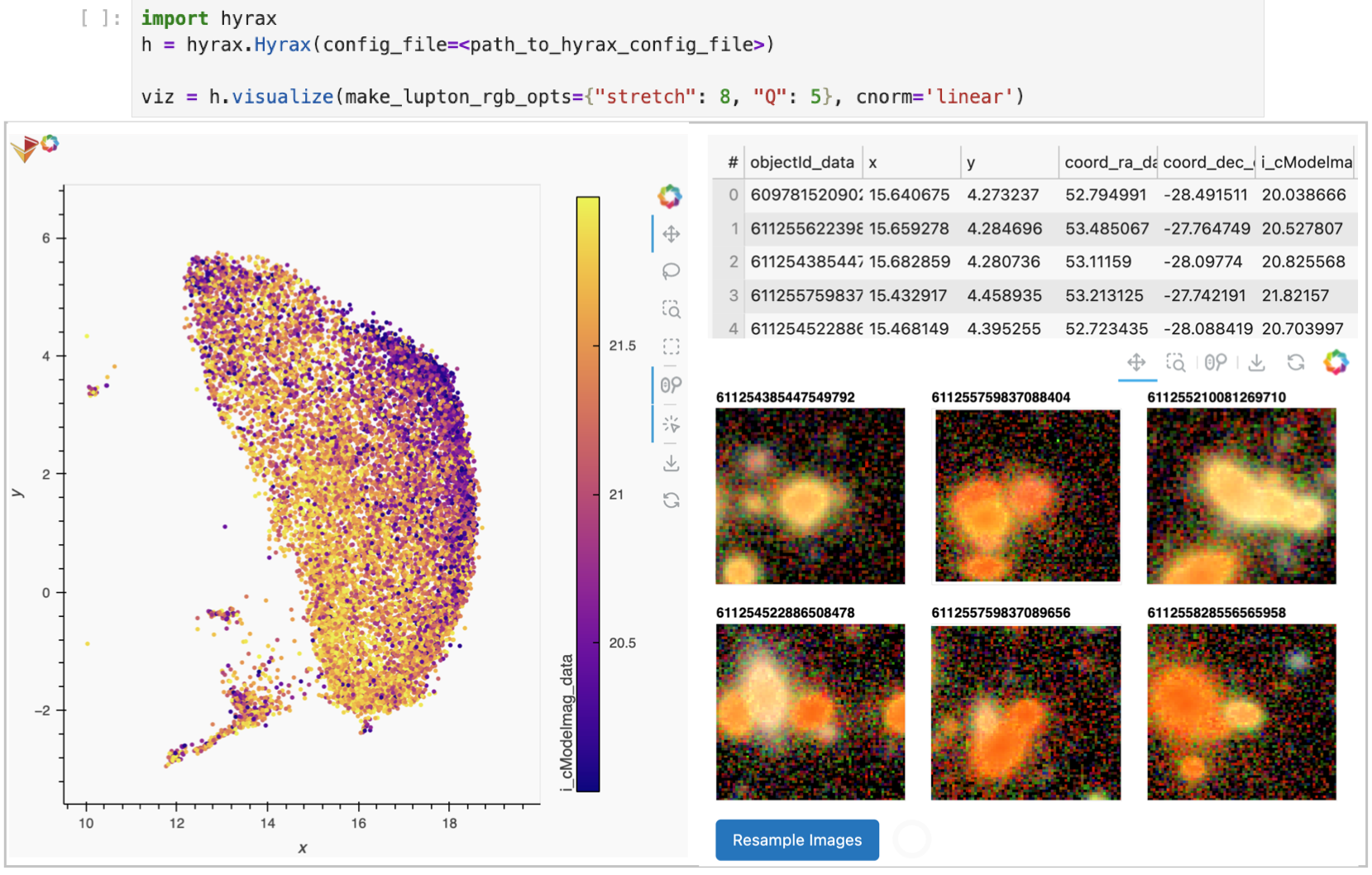}
    \end{interactive}
    \caption{Two-dimensional latent-space visualization produced using the Hyrax 2D Latent Space Explorer. The animated version of this figure (available in the online version and at \href{https://youtu.be/KSQdVb73Wb0}{this link}) demonstrates interactive exploration of the latent space, allowing users to color objects by catalog-level properties and to select regions of the embedding space for further inspection. The static snapshot also shows the \texttt{hyrax.visualize()} command used to generate the visualization, underscoring that the interactive tool runs entirely within a Jupyter Notebook environment. Selecting regions within the latent space triggers the retrieval and display of associated catalog information and imaging data for the selected points. This enables rapid inspection of latent-space structure and its correspondence with underlying observational properties, as well as qualitative assessment of whether structure in the embedding reflects meaningful astrophysical variation or survey-specific artifacts. While the above example shows imaging data, \texttt{hyrax.visualize()} can display any data source the Hyrax \texttt{DataProvider} (Figure~\ref{fig:hyraxql}) has access to---such as spectra or light curves---for objects selected in the latent space.}
    \label{fig:2d_viz}
\end{figure*}

Unsupervised and self-supervised models, such as autoencoders and representation-learning approaches based on contrastive objectives, can be trained relatively easily using well-defined loss functions. Yet their suitability for unsupervised discovery, clustering, or anomaly detection is often difficult to assess from scalar loss metrics alone. In practice, models with similar training losses can yield qualitatively different latent representations, with substantially different implications for scientific interpretability and downstream analysis. 

For astronomical datasets---where each data point corresponds to a complex astrophysical source with multi-dimensional observational properties---direct access to learned latent-space representations is indispensable. The ability to rapidly inspect objects occupying specific regions of latent space, including access to imaging, spectral, or time-series data together with associated catalog-level measurements, is critical for choosing between competing unsupervised models. Such interactive inspection also enables rapid assessment of whether apparent clustering or outliers in latent space correspond to scientifically interesting populations or are instead driven by instrumentation or survey-specific systematics. 

To date, astronomy lacks a general-purpose, survey-agnostic visualization framework for interrogating latent spaces produced by unsupervised models across surveys or heterogeneous data types. While some survey-specific tools exist \citep[e.g.,][]{Reis2021EffectivelySurveys, Storey-Fisher2021AnomalyNetworks}, they are typically tailored to specific datasets, object classes, or science cases. Hyrax addresses this gap by providing multiple visualization toolkits that operate uniformly on latent representations independent of survey origin and are tightly integrated with downstream analysis components (e.g., clustering). Hyrax's two-dimensional latent-space explorer is deployable within a Jupyter Notebook environment, while the three-dimensional latent-space explorer is implemented as a browser-based interactive interface. Both tools operate on latent embeddings produced during model inference, together with persistent object identifiers and associated metadata, after dimensionality reduction from the model’s latent space to two or three dimensions. Dimensionality reduction is currently performed using Uniform Manifold Approximation and Projection \citep[UMAP; ][]{McInnes2020UMAP:Reduction}, with planned support for additional methods such as t-distributed Stochastic Neighbor Embedding \citep[t-SNE; ][]{vanderMaaten2008tSNE}.

\begin{figure*}[htbp]
    \begin{interactive}{animation}{3d_viz_animated_ffmped_compressed.mp4}
    \centering
    \includegraphics[width=0.9\linewidth]{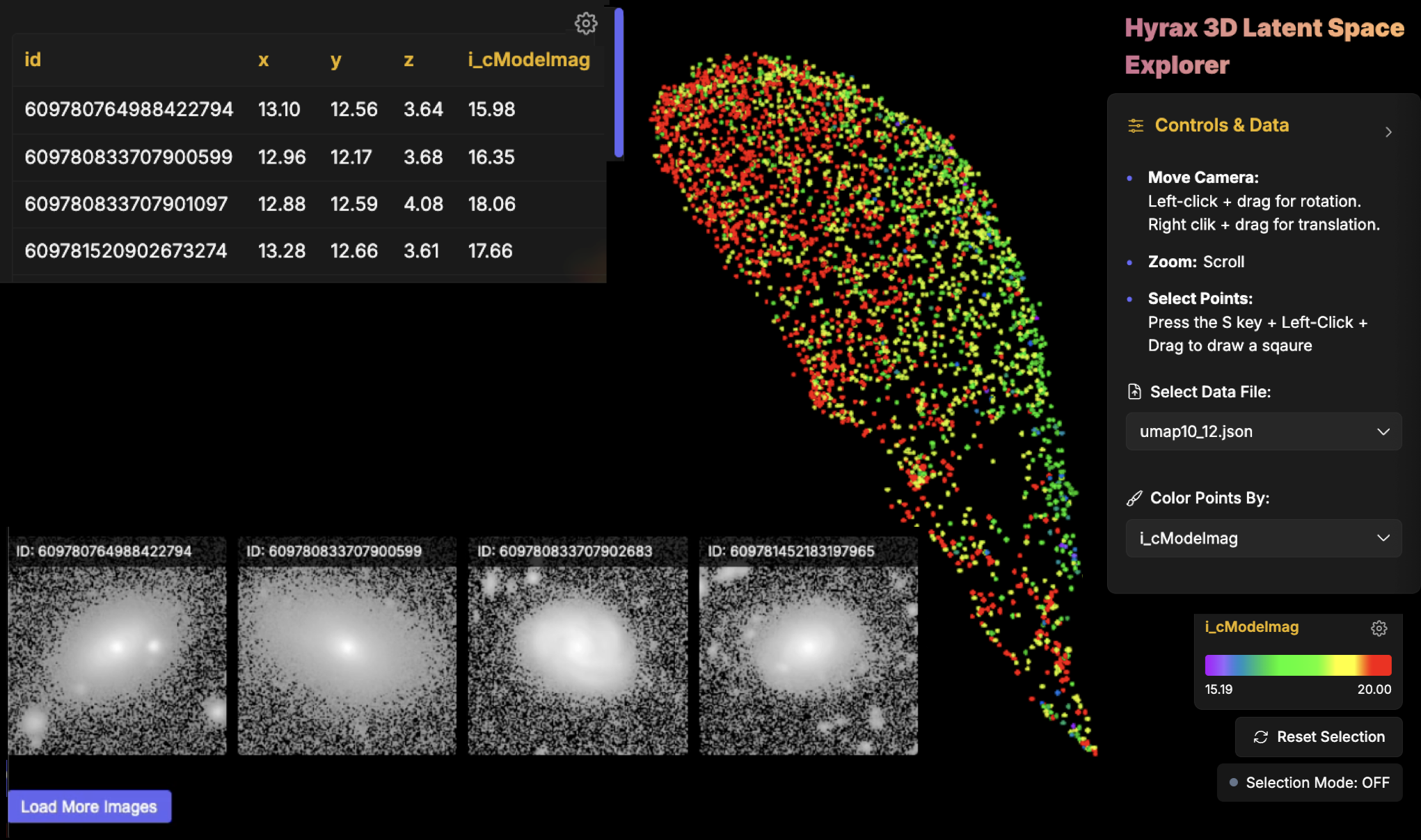}
    \end{interactive}
    \caption{Hyrax's browser-based 3D Latent Space Explorer allows users to explore three-dimensional representations of learned latent spaces interactively. The static image shown here is a snapshot from a video figure, and the video is available in the online version of the article and at \href{https://youtu.be/OqN6fQmpUg4}{this link}. Readers can also try out a live interactive demonstration at \href{https://hyrax-3d-demo.ghosharitra.com/}{this link}. The interface allows users to rotate, pan, and zoom the three-dimensional embedding to examine structures that may not be apparent in two-dimensional projections. As in the two-dimensional visualization tool (Figure \ref{fig:2d_viz}), selecting regions of the latent space retrieves the corresponding objects along with their associated catalog metadata and imaging data. Extensions to display spectral and time-series data for selected objects (when available) are currently under development.}
    \label{fig:3d_viz}
\end{figure*}

The visualization of the two-dimensional latent space can be accessed by users using a simple \texttt{hyrax.visualize()} verb as outlined in Figure \ref{fig:2d_viz}. The figure shows a representative snapshot (from a video) in which the latent-space representation of each object is projected into two dimensions. Hyrax provides users with the option to rasterize the point-cloud display to maintain interactive performance for large datasets. The interface is built on HoloViews with Datashader-enabled rendering, allowing scalable, interactive exploration directly within a Jupyter Notebook environment. Points may be colored by arbitrary scalar quantities, and users can interactively select regions of interest within the latent space. 

Internally, such selections yield the object identifiers occupying that region of the latent space. Hyrax then retrieves the corresponding catalog-level information for these objects, including any cross-matched properties provided by the user, and displays this information as an interactive table within the visualization interface. In parallel, the toolkit displays imaging data for a randomly selected subset of six objects drawn from the selection. Users may step through the full selection by resampling the displayed images using the provided controls. While the example in Figure~\ref{fig:2d_viz} shows imaging data, the display is not limited to images: because \texttt{hyrax.visualize()} is coupled directly to the Hyrax \texttt{DataProvider}, it can retrieve and display any data source the \texttt{DataProvider} has access to---such as spectra or light curves---for objects selected in the latent space, making the visualization interface directly aware of the full multimodal context available for each object. The selected data can also be exported as a Pandas \citep[][]{pandas} \texttt{DataFrame} within the same notebook for further inspection or passed directly to downstream analysis steps, such as clustering or similarity search. 

Three-dimensional latent-space visualization is provided through a browser-based interactive interface, as illustrated in Figure~\ref{fig:3d_viz}. The figure shows a representative snapshot extracted from an interactive video demonstrating exploration of a three-dimensional latent space. The interface enables users to rotate, pan, and zoom the latent-space projection, providing additional flexibility for exploring complex structure that may be compressed or ambiguous in two-dimensional views, particularly in dense regions of the embedding. As with the two-dimensional visualization described above, users can interactively select regions of latent space, with such selections yielding the corresponding set of object identifiers and dynamically retrieving associated catalog-level metadata and imaging data. The three-dimensional interface is built using the \texttt{three.js} library, enabling responsive, GPU-accelerated interaction directly within a standard web browser.

The two visualization modes serve complementary roles within the Hyrax workflow. The two-dimensional interface emphasizes tight integration with downstream analysis and programmatic workflows and, because it is deployable within a Jupyter Notebook, can be used directly on modern survey computing platforms such as the Rubin Science Platform and the Roman Research Nexus. The three-dimensional interface, by contrast, prioritizes intuitive spatial exploration of complex latent representations. Together, these tools support an iterative unsupervised discovery workflow in which users identify structures in latent space, retrieve related objects via similarity search or clustering, and assess their scientific coherence through direct inspection of the underlying data.

\section{Applying Hyrax to Different Astronomical Challenges} \label{sec:applications}

The flexible and modular architecture of Hyrax, described in \S\ref{sec:hyrax_description}, is designed to support ML workflows in a wide range of astronomical domains, data types, and learning paradigms. In this section, we present five representative applications that together illustrate this breadth: unsupervised and self-supervised representation learning applied to galaxy images in Rubin data (\S\ref{sec:rubin_dp1_lsst}); a hybrid approach combining unsupervised dimensionality reduction with interactive visual inspection for the discovery of cluster-scale gravitational lenses (\S\ref{sec:cluster_lenses}); supervised multimodal classification of transients from heterogeneous data streams including light curves, spectra, image cutouts, and metadata (\S\ref{sec:applecider}); supervised filtering of moving-object candidates in shift-and-stack searches for distant solar system objects (\S\ref{sec:kbmod}); and supervised detection of semi-resolved dwarf galaxies in HSC and LSST-like imaging using synthetic source injection enabled by the Rubin Science Pipelines (\S\ref{sec:chrysomallos}). Each application operates on real survey data; the results presented here are intended as first science previews that motivate dedicated follow-up publications rather than as final complete studies.

\subsection{Unsupervised Discovery \& Representation Learning in Rubin Data Preview 1} \label{sec:rubin_dp1_lsst}

The combination of depth, angular resolution, and sky coverage of upcoming surveys, illustrated in Figure~\ref{fig:data_growth}, will extend our census of extragalactic systems into regions of physical parameter space---stellar mass, surface brightness, size, and morphology---that remain sparsely sampled by current surveys. This expansion dramatically increases the likelihood of encountering rare or anomalous systems, but only if our analysis methodologies are capable of reliably isolating them from the ``regular" objects and from instrumental artifacts. Historically, many consequential discoveries in extragalactic astronomy have emerged precisely from identifying systems that deviate from prevailing expectations: rare morphological classes, extreme emission-line galaxies, unusual gravitational configurations, or strongly disturbed structures.

Unsupervised and self-supervised representation learning provide a complementary strategy to supervised methods, which are inherently bounded by the taxonomy encoded in their training data. Unsupervised and self-supervised learning organize objects according to intrinsic correlations in feature space rather than externally imposed categories. These approaches are particularly well-suited for extragalactic populations that are intrinsically continuous or occupy transitional regimes between conventional classes: galaxy interactions unfold across a sequence of dynamical stages, diffuse tidal structures appear with varying prominence and geometry, and composite systems often straddle the boundaries of established labels. By learning latent embeddings directly from the data, these methods enable the identification of both emergent clusters and low-density regions in the embedding space that may correspond to rare systems, transitional configurations, or morphologically complex populations, without requiring labeled training examples.

\begin{figure*}[htbp]
    \centering
    \includegraphics[width=0.8\linewidth]{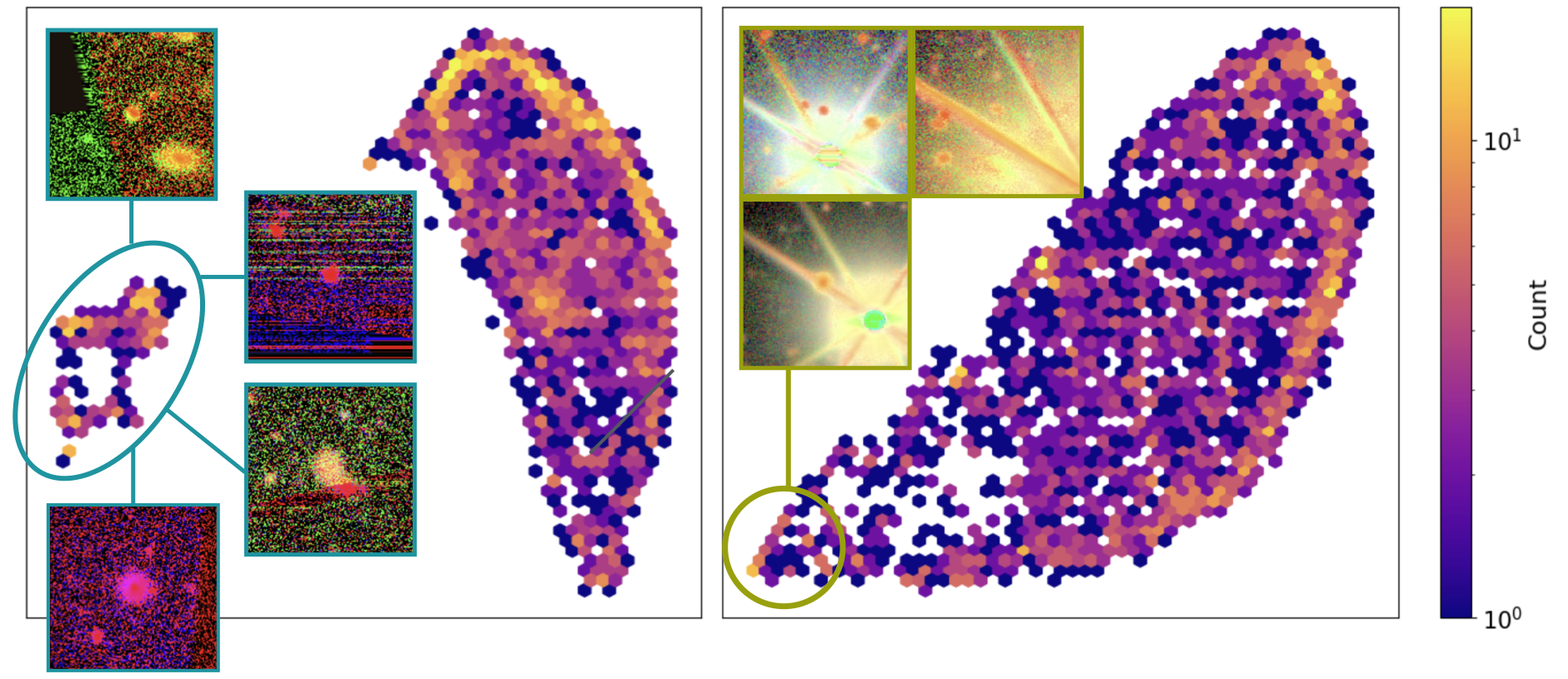}
    \caption{Two latent-space representations for the $i<20$ Rubin DP1 sources in ECDFS. (\textit{Left:}) Filter-specific artifacts/defects naturally segregate into a distinct island. (\textit{Right:}) Removing these and retraining reveals a second generation of outliers, now comprising imaging artifacts such as diffraction spikes and satellite trails/cosmic-ray hits. Each removal round progressively drives the sample toward a cleaner set of scientifically interesting extragalactic sources---all without any labeled training data.}
    \label{fig:artifacts_island}
\end{figure*}

\subsubsection{Rubin DP1 ECDFS Dataset}
To evaluate this approach on real survey data, we constructed latent-space embeddings using imaging data from Rubin-LSST Data Preview 1 \citep[DP1;][]{dp1_data}, focusing on the Extended Chandra Deep Field South (ECDFS) region of DP1, one of the commissioning fields. The DP1 dataset was obtained using the LSST Commissioning Camera during early on-sky operations and provides the first opportunity to implement and assess unsupervised discovery techniques on Rubin-quality coadds with realistic noise properties, multi-band coverage, and pipeline-processed data products. Within DP1, ECDFS offers a particularly suitable extragalactic testbed, with nearly uniform six-band ($ugrizy$) coverage across the entire field, relatively low stellar crowding, and the deepest coadds in DP1 (e.g., median $5\sigma$ depths of $g\sim26.18$, $r\sim25.96$, $i\sim25.71$). With depths within $\sim1$ mag of the anticipated 10-year LSST depth, ECDFS provides near-survey-end imaging quality within the controlled scope of a commissioning dataset. Additionally, the delivered image quality (median $i$-band PSF FWHM $\sim1\arcsec$) enables robust characterization of galaxy structure across a broad dynamic range in size and brightness. While the limited area reduces the likelihood of uncovering entirely new populations, it provides a practical environment for developing and testing the methodology on Rubin data before applying it to future wide-area releases.

From ECDFS, we selected extended sources by requiring \texttt{extendedness = 1} in at least one band, and further required that the source have imaging available in all six bands. This selection yielded $4\times10^5$ objects. To construct latent spaces for these objects, we trained a suite of unsupervised and self-supervised models within Hyrax. Specifically, we employed three different representation-learning architectures---a convolutional autoencoder \citep[][]{Hinton2006ReducingNetworks}, Simple Framework for Contrastive Learning of Visual Representations \citep[SimCLR; ][]{Chen2020ARepresentations}, and Bootstrap Your Own Latent \citep[BYOL; ][]{Grill2020BootstrapLearning}---spanning both reconstruction-based and contrastive/non-contrastive self-supervised learning paradigms. For each model, we explored a range of hyperparameter configurations, including the dimensionality of the latent embedding space, yielding several dozen distinct latent representations that capture different aspects of the underlying astronomical objects.

To examine the structure of the learned representations, we projected each latent space into two dimensions using the UMAP algorithm, as implemented within Hyrax. Thereafter, we explored these projections and the underlying higher-dimensional latent spaces using Hyrax's integrated tools for similarity search, clustering, and interactive latent-space exploration. We performed the above analysis separately on different magnitude-limited subsamples to restrict the range of physical scales and signal-to-noise regimes within each subsample. In \S \ref{subsec:dp1_results}, we report initial exploratory results for the $i<20$ and $20\leq i<22$ subsamples. A more comprehensive analysis of the full sample will be presented in an upcoming study (Ghosh et al., in prep.).

\subsubsection{Latent-Space Exploration Results} \label{subsec:dp1_results}
\begin{figure*}[htbp]
    \centering
    \includegraphics[width=1.0\linewidth]{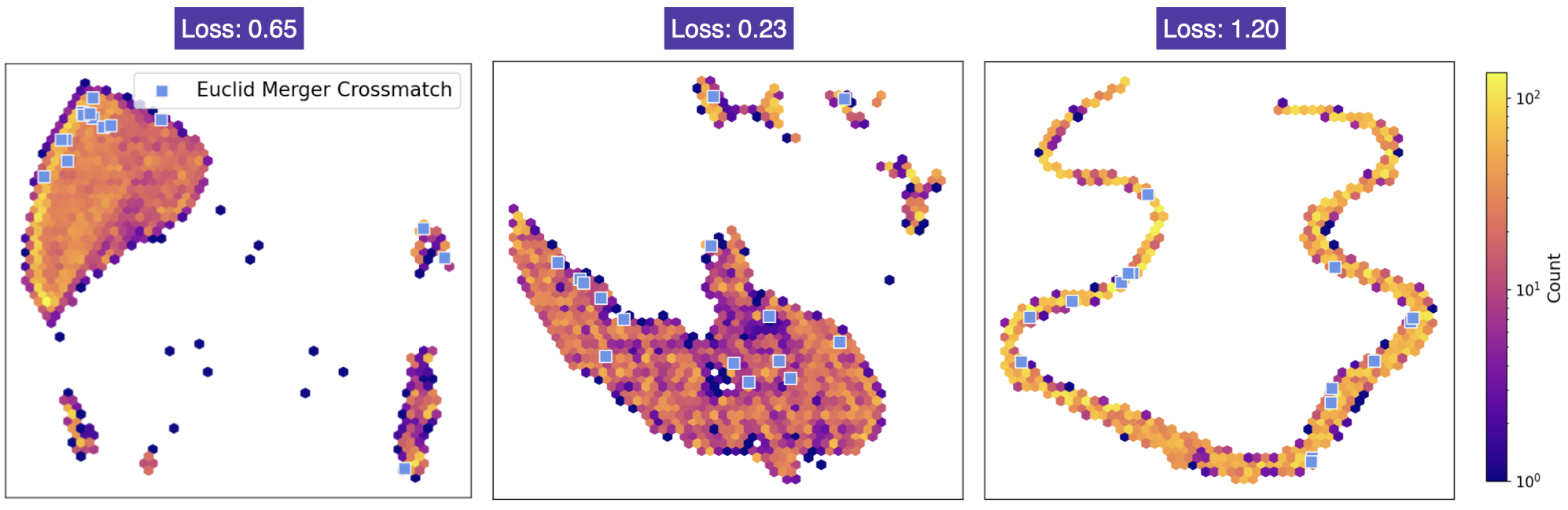}
    \caption{UMAP projections of latent spaces learned by three unsupervised models with varying hyperparameters, applied to $20\leq i<22$ ECDFS DP1 galaxies. Overlaid blue squares mark mergers identified in the Euclid Quick Release 1 visual morphology catalog \citep{EuclidCollaboration2025EuclidCatalogue}. The spatial clustering of the mergers varies non-monotonically with the validation loss obtained for each model (shown above each panel). This illustrates that training metrics alone are an unreliable proxy for the scientific utility of a learned representation, and underscores the value of Hyrax's visualization tools for model selection.}
    \label{fig:loss_clustering}
\end{figure*}
We begin by examining the global structure of the UMAP embeddings and show two such embeddings for the $i<20$ sample in Figure \ref{fig:artifacts_island}. As shown in the left panel, sources affected by filter-specific artifacts (e.g., cutouts with only partial coverage in one filter or those corrupted by filter-specific defects) naturally segregate into a distinct island, well-separated from the rest of the sample. Removing these sources and retraining the model reveals a second generation of outliers (shown in the right panel), comprising sources whose anomalous appearance arises not from filter-based corruption but from other imaging artifacts, such as diffraction spikes and satellite trails/cosmic-ray hits. This iterative peeling demonstrates that Hyrax can not only effectively flag survey artifacts and distinguish between qualitatively different classes of such artifacts but also do so in a fully unsupervised manner, without any labeled training data. The result is a progressively cleaner sample, with each removal round improving the scientific quality of the dataset. 

As discussed earlier, latent-space exploration is well-suited to quickly identify extragalactic populations that may be difficult to capture comprehensively with parametric selection or supervised algorithms alone. However, validating whether a learned representation has genuinely organized them into coherent structures is challenging in a new dataset such as Rubin DP1, where no pre-existing classification labels are available. Below, we describe a general pathway for addressing this using Hyrax's latent-space exploration and vector database tools.

First, we cross-matched our ECDFS DP1 sample against two external catalogs: i) the \citet{Tanoglidis2021ShadowsSurvey} catalog of low-surface-brightness (LSB) galaxies in the Dark Energy Survey Y3 Gold Catalog \citep{Sevilla-Noarbe2021DarkCosmology} (DES LSB catalog hereafter); ii) mergers identified in the \citet{EuclidCollaboration2025EuclidCatalogue} visual morphology catalog of galaxies in Euclid Quick Release 1 (Q1) \citep[][]{EuclidCollaboration2025EuclidOverview}. Cross-matching was performed using a greedy, $\chi^2$-based algorithm modeled on the LSST-Science pipeline's probabilistic matcher\footnote{\url{https://github.com/lsst/meas_astrom/blob/main/python/lsst/meas/astrom/matcher_probabilistic.py}}, which iterates through reference objects brightest-first and assigns each to the DP1 target with the lowest reduced $\chi^2$ computed from astrometric separation and multi-band photometry. From the Euclid catalog, we selected merging galaxies using the \texttt{merging\_merger} vote fraction with a threshold of $70\%$ (Euclid Merger Catalog hereafter). We then used Hyrax's latent-space visualizer to examine where these cross-matched objects lie within the learned latent-space representations and identified new candidates that occupy the same neighborhood. We also leveraged Hyrax's vector database to perform similarity searches using both the cross-matched objects and the newly visually identified candidates as queries, rapidly building samples of similar sources.

The cross-matching exercise reveals a subtle yet practically important observation, illustrated in Figure \ref{fig:loss_clustering}: the loss value obtained during training/validation may not be the best guide to the scientific quality of a learned representation. The three panels show UMAP projections of latent spaces produced by models with varying hyperparameters for the $20\leq i<22$ sample, with cross-matched Euclid mergers overlaid as blue reference squares. The spatial coherence of these mergers across the three embeddings does not track monotonically with the validation loss---the intermediate-loss model (left panel) produces a markedly more compact clustering. Without the ability to rapidly generate and interactively inspect such embeddings across many models, selecting the most scientifically useful representation from quantitative training metrics alone would be difficult. Hyrax's integrated tools make this kind of comparative assessment a natural and efficient part of the unsupervised discovery workflow.

\begin{figure*}[htbp]
    \centering
    \includegraphics[width=1.0\linewidth]{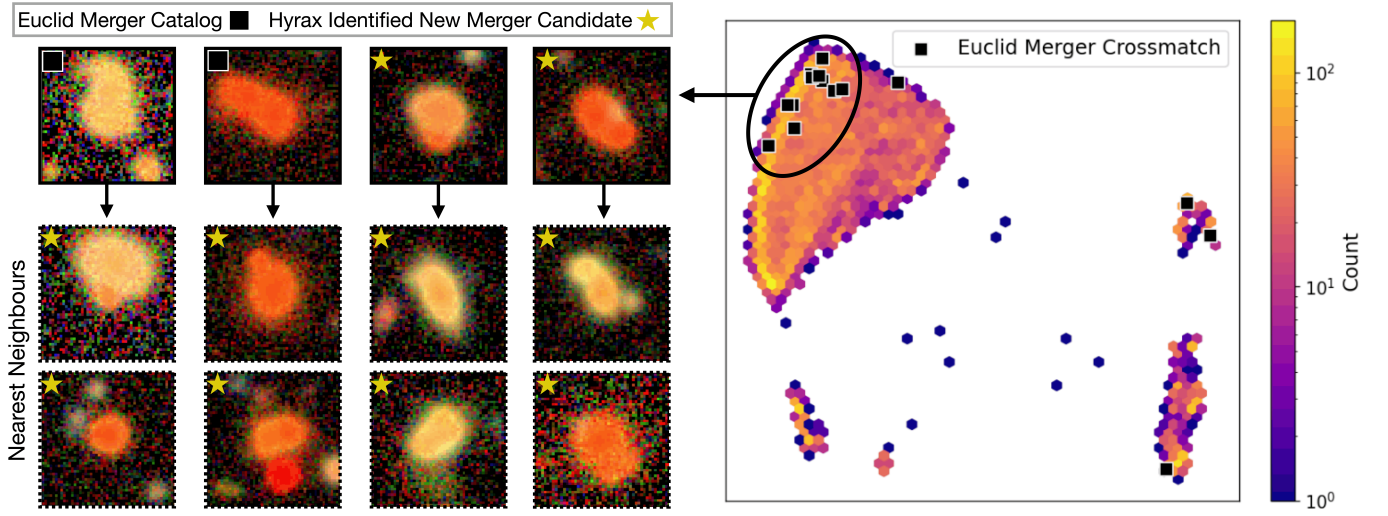}
    \caption{\textit{(Right:)} The UMAP that corresponds to the strongest clustering of merger candidates for the $20\leq i<22$ Rubin DP1 ECDFS sample is shown here, with Euclid merger catalog cross-matches denoted by black squares. We explore the region denoted by the black ellipse using Hyrax's 2D Latent Space Explorer. \textit{(Left:)} The top row shows example objects from this region: two randomly selected Euclid-catalog mergers (black squares) and two randomly selected new merger candidates that we find (gold stars). To demonstrate that this truly represents a neighborhood of merging/interacting galaxies, we show the nearest neighbors of each galaxy in the bottom two rows. Almost all the retrieved neighbors, although not present in the Euclid merger catalog, show clear signatures of ongoing or recent major/minor mergers. This demonstrates how Hyrax can be used to rapidly assemble large, diverse samples of merging systems from new survey data, without requiring any labeled training examples.}
    \label{fig:euclid_mergers}
\end{figure*}

\begin{figure*}[htbp]
    \centering
    \includegraphics[width=0.9\linewidth]{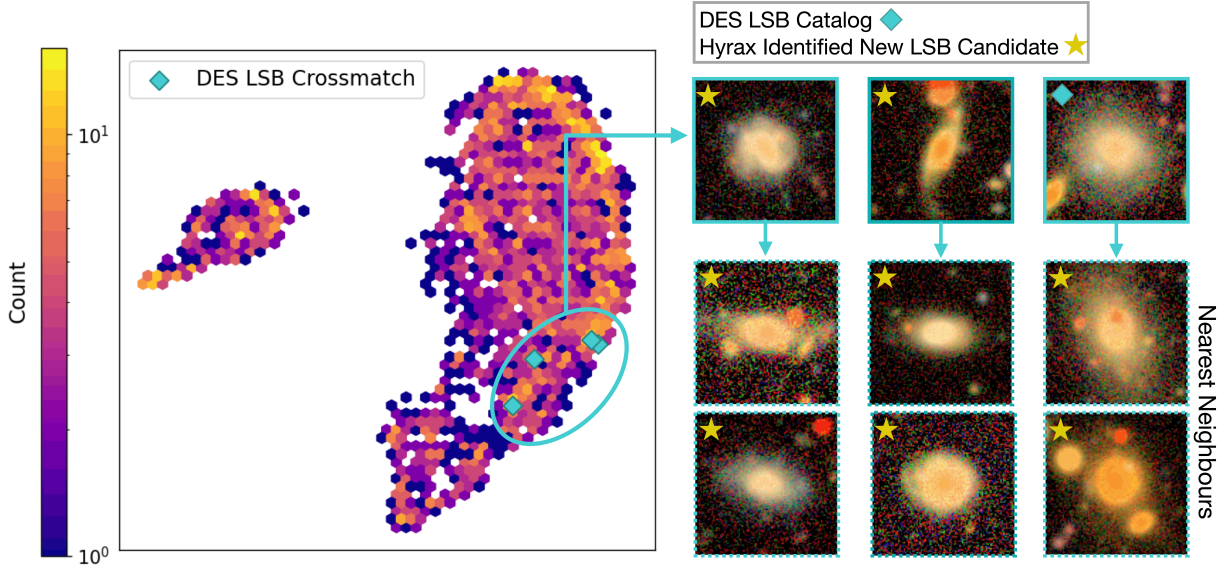}
    \caption{\textit{(Left:)} A UMAP projection of the best-performing latent-space representation for the $i<20$ Rubin DP1 ECDFS sample is shown here, with cross-matched DES LSB catalog objects marked as cyan diamonds. The cross-matched objects are compactly clustered within the cyan ellipse, which we investigate further using Hyrax's 2D latent space explorer. \textit{(Right:)} The top row shows one randomly selected DES LSB catalog match (cyan diamond) and two randomly chosen new LSB candidates (gold stars) that we find in this region. The bottom two rows show the nearest neighbors for each source in the top row. Almost all retrieved neighbors display diffuse, extended morphologies characteristic of LSB systems, despite their absence from the DES LSB catalog, further confirming the coherence of this latent-space neighborhood.}
    \label{fig:des_lsb}
\end{figure*}

We examine the best-performing representation from Figure \ref{fig:loss_clustering} in detail in Figure \ref{fig:euclid_mergers}. Using Hyrax's interactive 2D latent-space explorer, we inspected the compact region marked with a black ellipse in the right panel of Figure \ref{fig:euclid_mergers} and identified numerous additional merger candidates, encompassing both close-pair systems with clear tidal distortions and morphologically disturbed galaxies consistent with ongoing interactions at various dynamical stages. Two randomly chosen newly identified candidates, demarcated by gold stars, are shown in the first row of the left panel of Figure \ref{fig:euclid_mergers}. 

To demonstrate that these sources genuinely occupy a coherent merger neighborhood in the latent space, we retrieved the nearest neighbors of both newly identified candidates and of the Euclid-catalog mergers using Hyrax's vector database similarity search. Figure~\ref{fig:euclid_mergers} therefore also shows the nearest neighbors to the newly identified candidates, along with neighbors retrieved using two randomly selected Euclid-identified mergers as query objects. The retrieved neighbors---none of which appear in the Euclid merger catalog---consistently display morphological signatures of galaxy interactions, including tidal distortions, asymmetric light distributions, and close companion systems. This suggests that the representation shown has indeed organized galaxies with a variety of interaction-driven morphologies into a coherent region of the embedding space.

The fact that most of the merger candidates recovered in Figure \ref{fig:euclid_mergers} are absent from the Euclid merger catalog is noteworthy, given that Euclid Q1 covered the full DP1 ECDFS footprint. However, two considerations should be kept in mind. First, the Euclid catalog was constructed through visual classification, whereas the candidates here are identified through an entirely automated, data-driven process---the two approaches are therefore complementary rather than directly comparable. Second, our designation of these sources as ``new" refers strictly to their absence from the Euclid sample; a broader cross-match against other available classifications and imaging of ECDFS will be reported in Ghosh et al. (in prep.), where we will also provide a more rigorous assessment of completeness and purity. The examples shown here also suggest that the greater depth of Rubin-LSST imaging at full survey depth (despite its lower spatial resolution) enables the recovery of fainter or more diffuse interaction signatures that may fall below the detection threshold of single-epoch space-based data. More broadly, this illustrates how deep ground-based imaging can be used to rapidly identify large numbers of diverse interacting systems, which can subsequently be followed up at higher angular resolution with space-based facilities.

An analogous picture also emerges for low-surface-brightness objects, as shown in Figure \ref{fig:des_lsb}. The cross-matched DES sources with LSB features occupy a specific region of the latent space at the periphery of the main galaxy locus. Inspecting this region (marked with a cyan ellipse in the figure) with Hyrax's latent-space explorer, we identified several additional sources exhibiting similar low-surface-brightness features that were not present in the reference catalog. Two randomly selected examples of these newly identified candidates, along with one cross-matched DES LSB object, are shown in the first row of the right panel of Figure \ref{fig:des_lsb}.

To assess whether these objects genuinely occupy a coherent LSB neighborhood in the latent space, we retrieved their nearest neighbors via Hyrax's similarity search. The retrieved neighbors, shown in the bottom two rows, almost universally display the diffuse, extended light distributions characteristic of LSB systems---yet none appear in the DES LSB catalog. The absence of these sources from the reference catalog likely reflects a combination of the greater depth of Rubin DP1 imaging and the data-driven nature of the latent-space selection. A rigorous completeness assessment will be presented in Ghosh et al. (in prep.).

Together, Figures \ref{fig:euclid_mergers} and \ref{fig:des_lsb} demonstrate that a single unsupervised framework can simultaneously surface structurally distinct populations (interacting systems and diffuse LSB galaxies) in new survey data, organizing them into coherent and scientifically interpretable regions of a latent space, without requiring any labeled training examples. This capability will become increasingly valuable as Rubin DP2, Rubin DR1, and future releases dramatically expand the size of the dataset.

\subsection{Discovering Cluster-scale Gravitationally Lensed Arcs: Hyrax Visualization of Clustering and Latent Spaces} \label{sec:cluster_lenses}

Gravitational lensing provides a uniquely powerful probe of otherwise inaccessible physical scales in the high-redshift Universe, magnifying background sources beyond the resolving power of current facilities. A hallmark of recent ground-based imaging surveys has been their ability to build a systematically selected sample of gravitational lenses with high-redshift arcs behind them. The most magnified arcs of these reside in the field of view of galaxy groups and clusters, which are far sparser than their galaxy-galaxy lens system counterparts \citep[e.g.,][]{Bayliss2014, Khullar2021, Mork2025}. One of LSST's primary science goals for the field of extragalactic astronomy is to identify intermediate and low-brightness lensed arcs at $z_{\rm spec} > 1$, probing the physics of star formation, nebular gas, and dust in high-redshift galaxies through ground- and space-based spectroscopic follow-up. The first step in this process is to build a ground-based imaging sample. 

Existing lens-finding efforts employ a combination of supervised and unsupervised ML alongside visual inspection of targeted survey fields. In general, automated approaches offer greater scalability but can suffer in completeness or purity relative to careful human inspection, which remains the gold standard for cluster-scale arcs but is prohibitively time-consuming at survey scale \citep[e.g.,][]{Bayliss2014, Khullar2021, Sukay2022, Cloonan2025}. Hyrax addresses this tension directly: its latent-space visualization and similarity-search tools enable rapid, systematic inspection of candidate fields flagged by unsupervised clustering, combining the efficiency of automated selection with the precision of targeted visual review.

We demonstrate one such application here, in which Hyrax's built-in LSST Image Dataset Class (Figure~\ref{fig:lsst_downloaded_dataset}) is used to generate and manage image cutouts of candidate lensing fields identified through a hybrid approach combining unsupervised clustering and visual inspection. The hypothesis is that by finding overdensities of galaxies in the joint position-redshift space, we are likely to build a parent sample of fields of view that host faint or bright lensed arcs.

\begin{figure*}[htbp]
    \centering
    \includegraphics[width=1.0\linewidth]{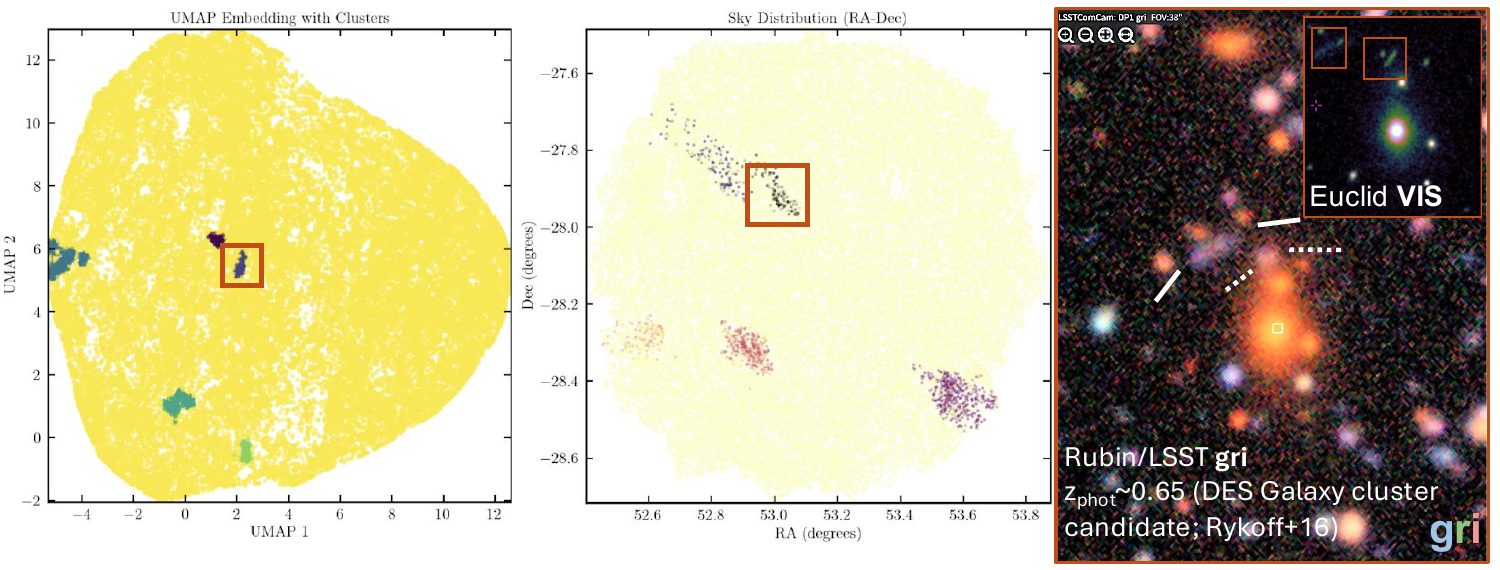}
    \caption{Example of our hybrid lens and arc finding approach. (Left and Center) UMAP and RA+Dec overdensities corresponding to groups and clusters in ECDFS, one of the fields sampled by Rubin DP1 observations. (Right) Post visual inspection, we isolate two lensed arc candidates (marked with solid and dotted lines) in the pan-chromatic Rubin $gri$ and Euclid VIS imaging (see inset).}
    \label{fig:lens}
\end{figure*}

In this hybrid approach, we utilize the UMAP module to construct a latent-space embedding of Rubin DP1 ECDFS galaxies with three inputs---Right Ascension, Declination, and photometric redshift z$_{\rm phot}$ from LePHARE \citep[][]{Arnouts1999LePhare, Ilbert2006LePhare} included in DP1 \citep[][]{Zhang2025RAIL_DP1}---all accessed via LSDB\footnote{\url{http://docs.lsdb.io/}} \citep[][]{Caplar2025LSDB, Malanchev2025LSDB_DP1}. Using the Density-Based Spatial Clustering of Applications with Noise algorithm \citep[DBSCAN;][]{Ester1996DBSCAN}---a ``cluster finding'' algorithm---we find overdensities in UMAP space that likely correspond to overdensities of galaxies in the joint RA-Dec-z$_{\rm phot}$ space. In these overdensities, we build 10" and 20" cutouts to facilitate visual inspection.

Through visual inspection, we isolate potential lensed arc candidates (without spectroscopic confirmation, these could very likely be edge-on galaxies, or systems with non-axisymmetric morphology due to mergers/outflows). In Figure \ref{fig:lens}, we demonstrate an example of our lens finding framework---our hybrid workflow isolates a lensed arc candidate in the field of a galaxy cluster in the ECDFS field (a Dark Energy Survey overdensity at z$_{\rm phot} \sim$0.65; \citealt{rykoff2016}). Upon cross-matching with Euclid VIS archival observations, the morphology of the arc is consistent with a sheared lensed arc (as well as the detection of a second candidate otherwise blended in Rubin $ugrizy$ observations). The details of spectroscopic follow-up of this system with Fast Turnaround Gemini GMOS observations will be discussed in a separate publication (Khullar et al., in prep.). 

\subsection{Multimodal Learning for ZTF Early Classification}
\label{sec:applecider}
As an example of how a complex, externally developed multimodal system can be deployed within Hyrax's modular infrastructure, we present the integration of \texttt{AppleCiDEr} \citep[Applying Multimodal Learning to Classify Transient Detections Early;][]{applecider}---a multimodal deep-learning framework designed for the early classification of astronomical transients in large time-domain surveys such as the Zwicky Transient Facility \citep[ZTF;][]{Bellm2019, Graham2019} and LSST. 

\begin{figure}[h]
  \centering
  \begin{minipage}[t]{0.85\columnwidth}
    \centering
    \includegraphics[width=\textwidth]{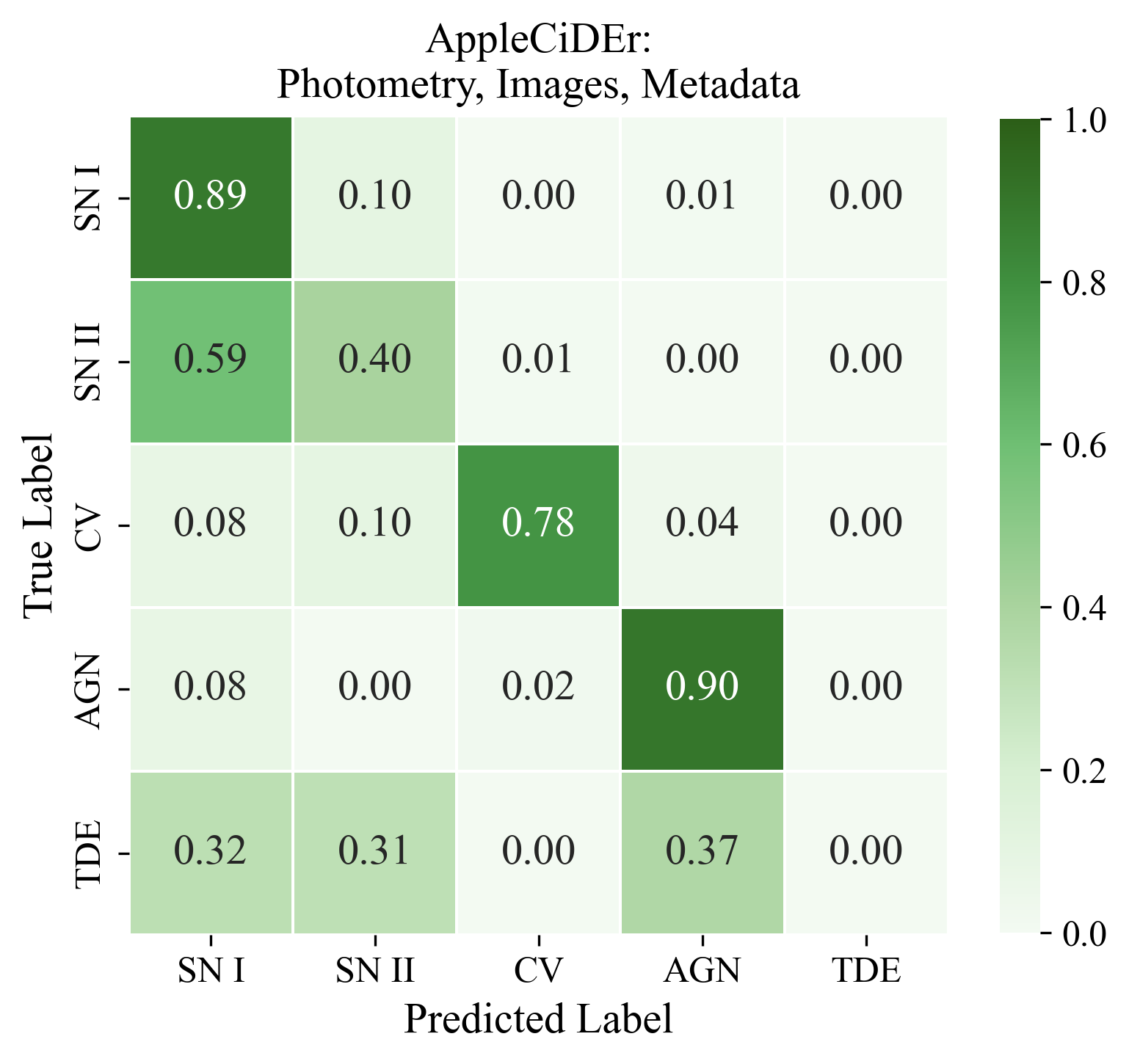}\\
    {\footnotesize (a) Three modalities: photometry, images, and metadata.}
  \end{minipage}\hfill
  \\
  \begin{minipage}[t]{0.85\columnwidth}
    \centering
    \includegraphics[width=\textwidth]{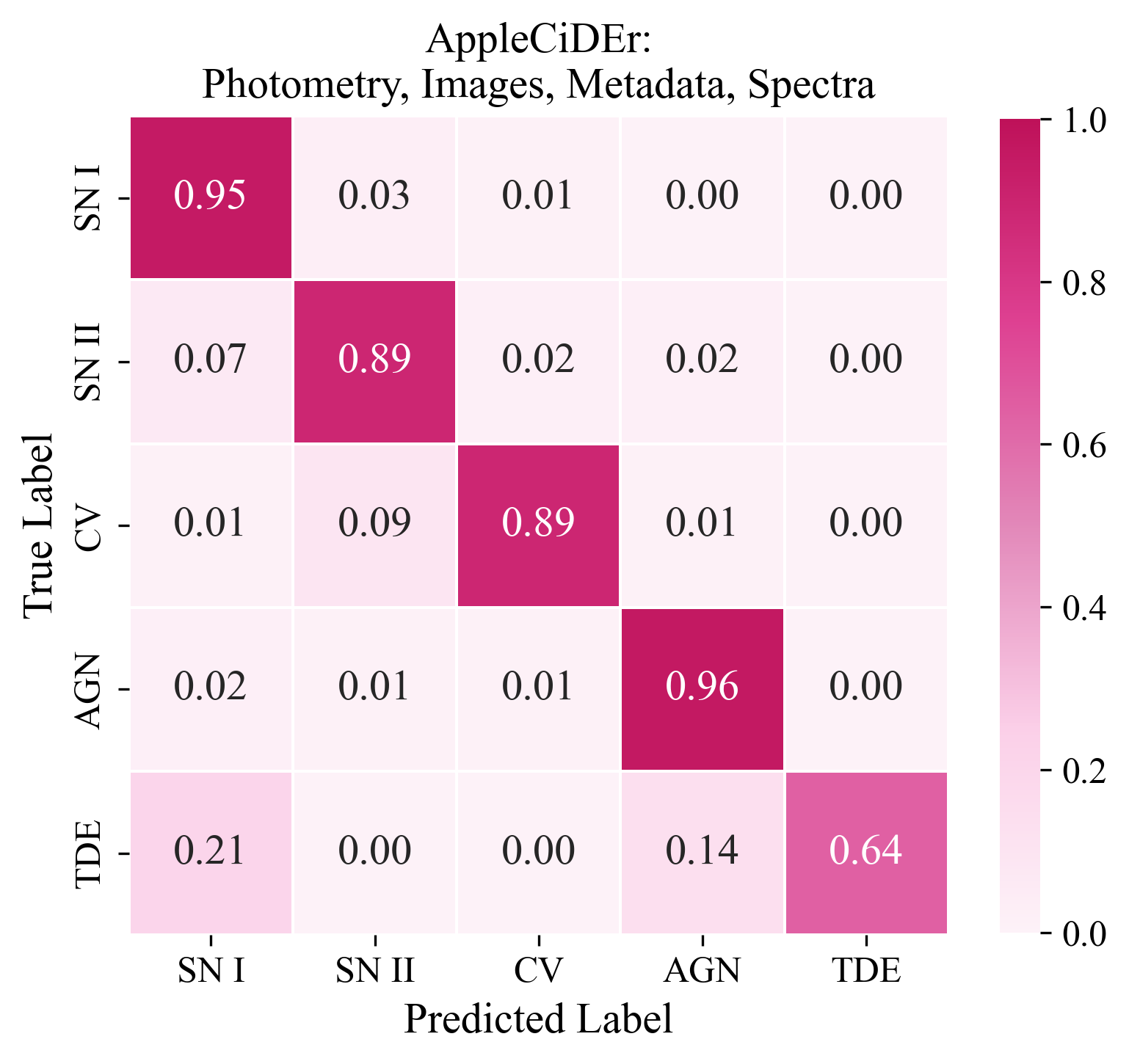}\\
    {\footnotesize (b) Four modalities: photometry, images, metadata, and spectra.}
  \end{minipage}
  \caption{Normalized confusion matrices of \texttt{AppleCiDEr} using \textit{(a)} three and \textit{(b)} four modalities. The classification of TDEs is more challenging due to the rarity of this class and the limited number of training samples. The comparison highlights the importance of multimodal training.}
  \label{fig:cider_confmats}
\end{figure}

\texttt{AppleCiDEr} integrates heterogeneous data streams, photometric light curves, spectra, image cutouts, and contextual metadata within a unified architecture to improve classification accuracy at early phases. Its design includes transformer encoders for photometric sequences, a convolutional neural network for spectral analysis \citep{appleciderII}, and a multimodal fusion module employing a Mixture-of-Experts strategy for images and metadata. By leveraging complementary information across modalities, \texttt{AppleCiDEr} achieves strong classification accuracy across all five transient classes, as quantified by the confusion matrices in Figure~\ref{fig:cider_confmats}.

Within Hyrax, \texttt{AppleCiDEr} is implemented using custom DatasetClass modules for heterogeneous inputs. Training and inference are managed via \texttt{hyrax.train()}, enabling automated configuration tracking. The framework is publicly available as open-source software\footnote{\url{https://github.com/skyportal/applecider}} and is designed for integration within alert-broker ecosystems such as \texttt{BOOM}~\citep{BOOM}, an alert processing and broker framework for transient surveys. By combining scalable ML infrastructure with operational survey workflows, \texttt{AppleCiDEr} represents a practical solution for automated classification in the data-intensive era of next-generation synoptic surveys.

To demonstrate \texttt{AppleCiDEr}'s performance, we show here (Figure~\ref{fig:cider_confmats}) the results of running \texttt{AppleCiDEr} on \textit{(a)} three modalities and \textit{(b)} four modalities. The dataset comprises approximately 30,000 real alerts and a total of 18,245 objects distributed across five classes: supernovae type I (SN~I) and II (SN~II), cataclysmic variables (CV), active galactic nuclei (AGN), and tidal disruption events (TDE). For each object, multiple data modalities are collected, including multi-band photometry (g, r, i), triplets of image cutouts (science, reference, difference), 17 selected metadata features, and spectra when available. Alert information is queried through the Kowalski broker \citep{duev2019}, while spectroscopic data are fetched from Fritz/SkyPortal \citep{vanderWalt2019, 2023ApJS..267...31C}. Preprocessing steps include magnitude-to-flux conversion, temporal merging of photometric measurements, feature normalization, and rest-frame correction of spectra with resampling onto a fixed grid. Full details of the preprocessing pipeline are provided in \citet{applecider}.

The inclusion of spectral information reduces class degeneracies that are difficult to disentangle from photometry alone, particularly between thermonuclear and core-collapse supernovae, and between TDEs and AGN. \texttt{AppleCiDEr} serves as a concrete demonstration of how Hyrax abstracts the surrounding ML infrastructure, allowing complex multimodal architectures to be deployed. 

\begin{figure*}[htbp]
  \centering
  \begin{minipage}{0.95\textwidth}
    \centering
    \includegraphics[width=\textwidth]{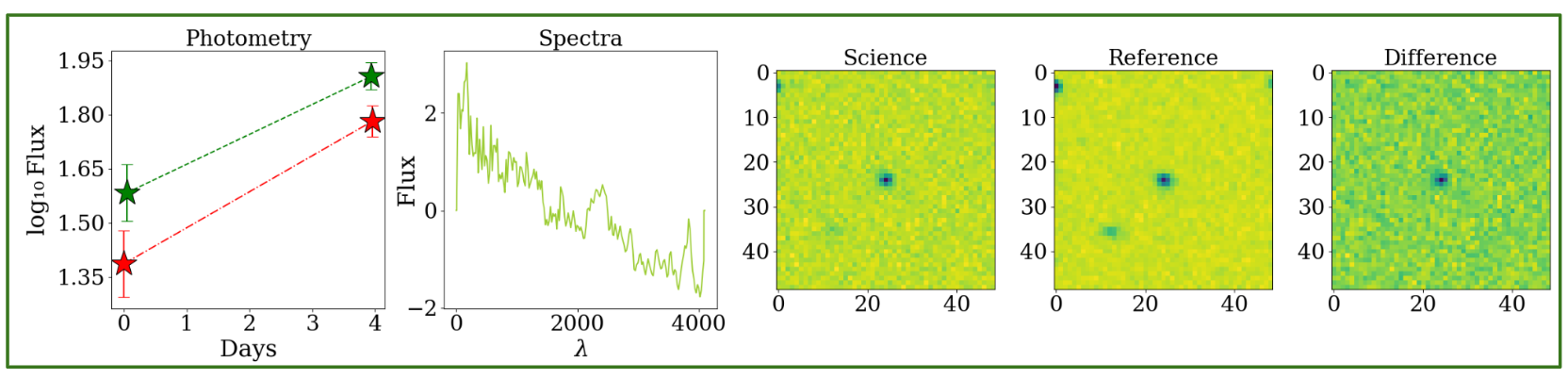}
    {\footnotesize (a) Misclassified TDE as SN II.}
  \end{minipage}
  \vspace{0.4cm}
  \begin{minipage}{0.95\textwidth}
    \centering
    \includegraphics[width=\textwidth]{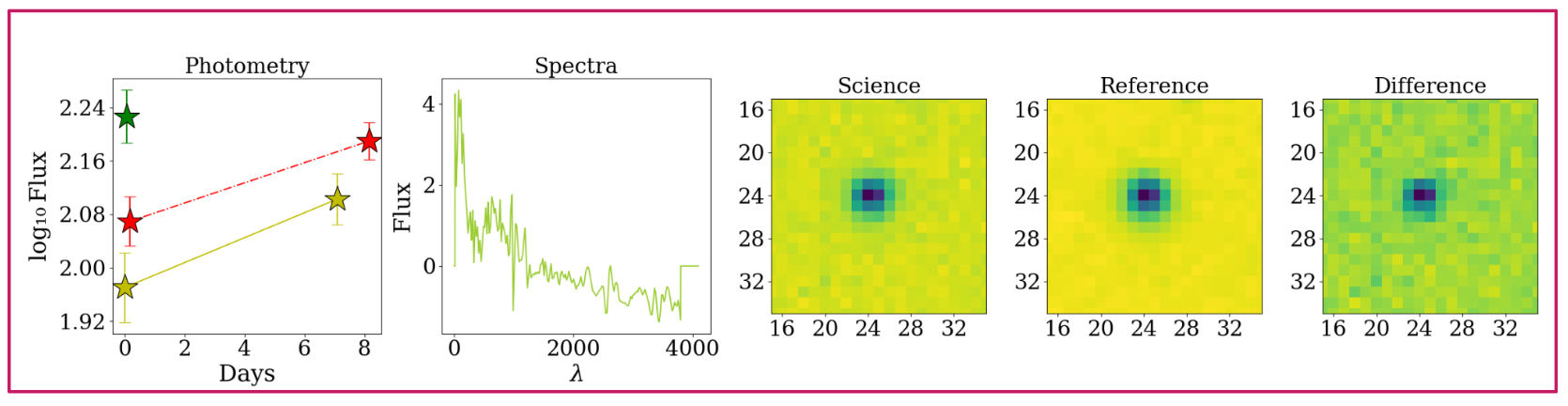}
    {\footnotesize (b) Correctly classified TDE.}
  \end{minipage}

  \caption{
  Multimodal visualization of two TDEs. 
  Panel (a) shows a TDE predicted as SN II. Panel (b) shows a correctly classified TDE where stronger spectral and photometric transient signatures are present. 
  The comparison illustrates the challenges of early-time multimodal classification, where incomplete photometric evolution and limited spectral information can lead to confusion between different classes of optical transients.
  }
  \label{fig:tde_case_study}
\end{figure*}

The comparatively lower performance on TDE classification is primarily driven by the extreme class imbalance, with only 54 TDE-samples confirmed. This scarcity limits the model’s ability to learn robust decision boundaries for this rare class and highlights the broader challenge of supervised learning for intrinsically uncommon transients. Figure~\ref{fig:tde_case_study} presents a representative comparison between a misclassified and a correctly classified TDE. In the upper panel, a spectroscopically confirmed TDE is predicted as a Type II Supernova (SN II). In the model’s input features, the event exhibits relatively modest transient signatures in the difference images, limited early-time photometric sampling with gradual flux evolution, and a spectrum lacking strong high-ionization emission features typically associated with TDEs. These characteristics can phenomenologically resemble early phases of supernova explosions, leading the multimodal fusion module to assign a higher probability to the SN II class. The misclassification likely arises from the absence of strong spectral discriminants and from incomplete photometric evolution in the model inputs. This example illustrates the difficulty of distinguishing rare nuclear transients from other optical transients when only partial multimodal information is available.

In contrast, the lower panel of Figure~\ref{fig:tde_case_study} presents a TDE correctly classified by the model. In this case, the multimodal inputs provide stronger discrimination. The photometric sequence exhibits clearer transient evolution with significant multi-band flux variation, while the spectral encoder receives a pronounced blue continuum excess indicative of a thermal accretion-driven flare. The image cutouts also show a well-defined point-like transient in the difference image. Unlike the misclassified example, the combined modality embeddings in this event reinforce a coherent transient signature across image, temporal, and spectral encoders. This comparison illustrates how the availability of stronger spectral and temporal features enables the model to correctly identify TDEs despite potential overlap with other transient classes. 

More broadly, \texttt{AppleCiDEr}'s integration into Hyrax demonstrates that production-scale supervised multimodal pipelines---spanning heterogeneous data streams, custom dataset classes, and complex fusion architectures---can be deployed within a single unified framework without bespoke infrastructure for each new application.

\subsection{Searching for Distant Trans-Neptunian Objects}
\label{sec:kbmod}


\begin{figure*}[htbp]
    \centering
    \includegraphics[width=0.9\linewidth]{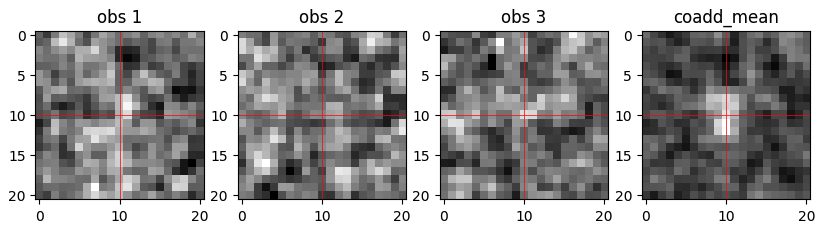}
    \caption{Observations of an inserted synthetic object. In each individual frame, the object is below the detection threshold, but in the coaddition centered on the object's trajectory, it is clearly visible.}
    \label{fig:ta228}
\end{figure*}

The positions and velocities of asteroids and small planetary bodies trace the history of the formation and evolution of the Solar System. For example, in the outer Solar System, Trans-Neptunian Objects (TNOs) contain dynamically-unperturbed relics from the formation of the Solar System \citep{Luu+Jewitt2002}. Their orbits reveal the history of the migration of massive planets from the inner to outer Solar System shortly after its formation \citep{NiceModel}. Improving our understanding of the sizes and orbital distributions of TNOs, especially at the low-mass end where they are poorly constrained, will be essential for testing these and other formation hypotheses.

\texttt{KBMOD} (Kernel-Based Moving Object Detection) is a digital tracking or shift-and-stack package that detects moving objects too faint to be seen in a single exposure \citep{Whidden_2019, Smotherman_2021}. It accomplishes this by searching along candidate trajectories within a stack of images, identifying and coadding the pixel data (Figure~\ref{fig:ta228}), filtering false positives, and visualizing the resulting candidate sources. Applying it to one month of Rubin data should increase the sensitivity of LSST to the detection of TNOs by a factor of 5 \citep{juric2019}.

\begin{figure}[htbp]
    \centering
    \includegraphics[width=0.9\linewidth]{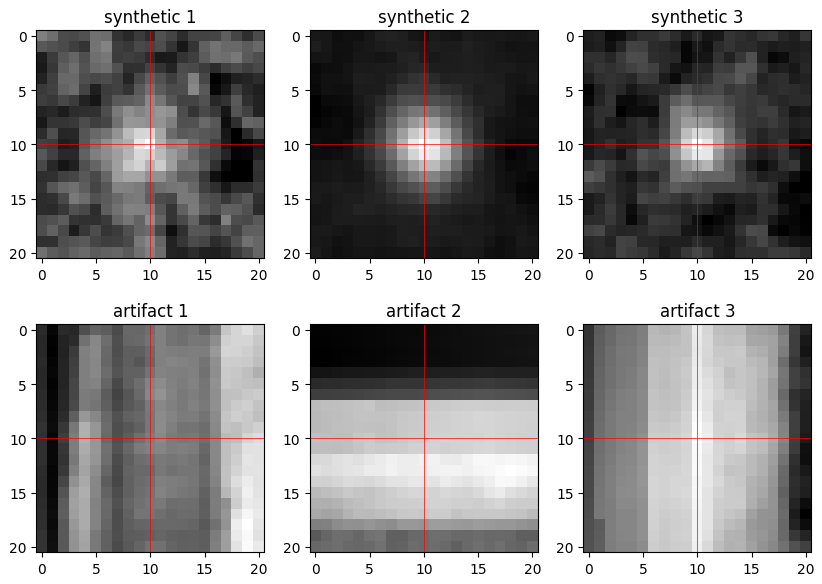}
    \caption{Coadded \texttt{KBMOD} result stamps, where the top row represents ``true" positive results (i.e., plausible moving objects) and the bottom row represents ``false" positive results (caused by imaging/differencing artifacts or unmasked static objects like galaxies).}
    \label{fig:kbmlstamp}
\end{figure}

\texttt{KBMOD} uses GPU acceleration to search billions of linear trajectories across a stack of observations and has been successful in discovering new TNOs as well as recovering known ones \citep{Smotherman_2021}. Due to the exhaustive nature of the core \texttt{KBMOD} algorithm it has, however, a very high false-positive rate (between 10:1 and 20:1, depending on the filtering steps used). The distinction between true and false positive stamps is fairly easy to determine visually (see Figure~\ref{fig:kbmlstamp}), which makes it a good candidate for an ML approach, specifically using a ResNet \citep{He:2016} model trained on \texttt{KBMOD} result stamps. 

%

In its initial implementation, we built a custom TensorFlow binary classifier distinguishing true and false positives. We recently adopted Hyrax as our general-purpose training and inference system, with significant improvements to our development workflow. We created a custom \texttt{KbmodStamps} dataset that reads in our results' coadded stamps and performs the training. The coadd stamps were derived by coadding a selection of pixels from each image along the pixel-space trajectory of the \texttt{KBMOD} tracklet and taking their mean. We used data from the Dark Energy Camera Ecliptic Exploration Project \citep[DEEP;][]{Trilling2024DEEP} survey, and the stamp size used was 21 by 21 pixels. For the true positive set of stamps, \texttt{KBMOD} results were matched to known inserted synthetic objects, and a subset of those results was taken as the training and test data for the ``true" label (meaning good objects that we would like to pass along from filtering). For the false positive set in DEEP, the \texttt{KBMOD} search parameters were set to look for objects traveling along trajectories of 85 degrees away from the solar system ecliptic (searching an orbit space that is highly unlikely to be real) to get a large set of imaging artifacts to use as training and test data for the ``false" label. For both training sets, there was also a series of augmentations to the stamps by mirroring, rotating, and shifting the origin by a pixel. The pre-augmentation true positive dataset contained ${\sim}9{,}000$ inserted synthetics, and the total size of the dataset was increased to ${\sim}100{,}000$ by augmentations. For the false positive, there were ${\sim}50{,}000$ imaging artifact stamps initially that grew to ${\sim}200{,}000$ after the augmentations.

For the first processing runs of DEEP data (with ${\sim}300{,}000$ tracklets left after the initial filtering steps), the model successfully filtered out 72.2\% of the total dataset for a final set of ``good" results of 87,490. Within that final set, 95.2\% of the inserted synthetic objects expected to pass the filter were recovered---demonstrating an ability to recover an overwhelming majority of the inserted synthetic sources. 

The switch to Hyrax enabled us to make, track, and deploy hyperparameter changes more quickly than before, with Hyrax's configuration-provenance system and TensorBoard integration providing a structured record of parameter changes across training runs. It also allowed us to more easily bring the trained models into our production pipelines for \texttt{KBMOD} processing. The \texttt{KBMOD} integration demonstrates how Hyrax's supervised learning infrastructure (dataset classes, configuration tracking, and experiment comparison tools) can be adopted into an existing production pipeline with minimal overhead, accelerating the development cycle for domain-specific classification tasks. Critically, this transition did not require us to independently learn or configure each of the underlying tools that Hyrax builds upon, such as MLflow for experiment tracking or TensorBoard for monitoring. Hyrax's unified interface absorbs that complexity, so the overhead of assembling a robust ML stack does not fall on domain scientists who are primarily focused on their scientific problem rather than software infrastructure.


\subsection{Searches for Semi-resolved Dwarf Galaxies in HSC and LSST: Supervised Learning Using Synthetic Source Injection Enabled by the Rubin Science Pipelines}
\label{sec:chrysomallos}
\begin{figure*}[htbp]
    \centering
    \includegraphics[width=0.98\linewidth]{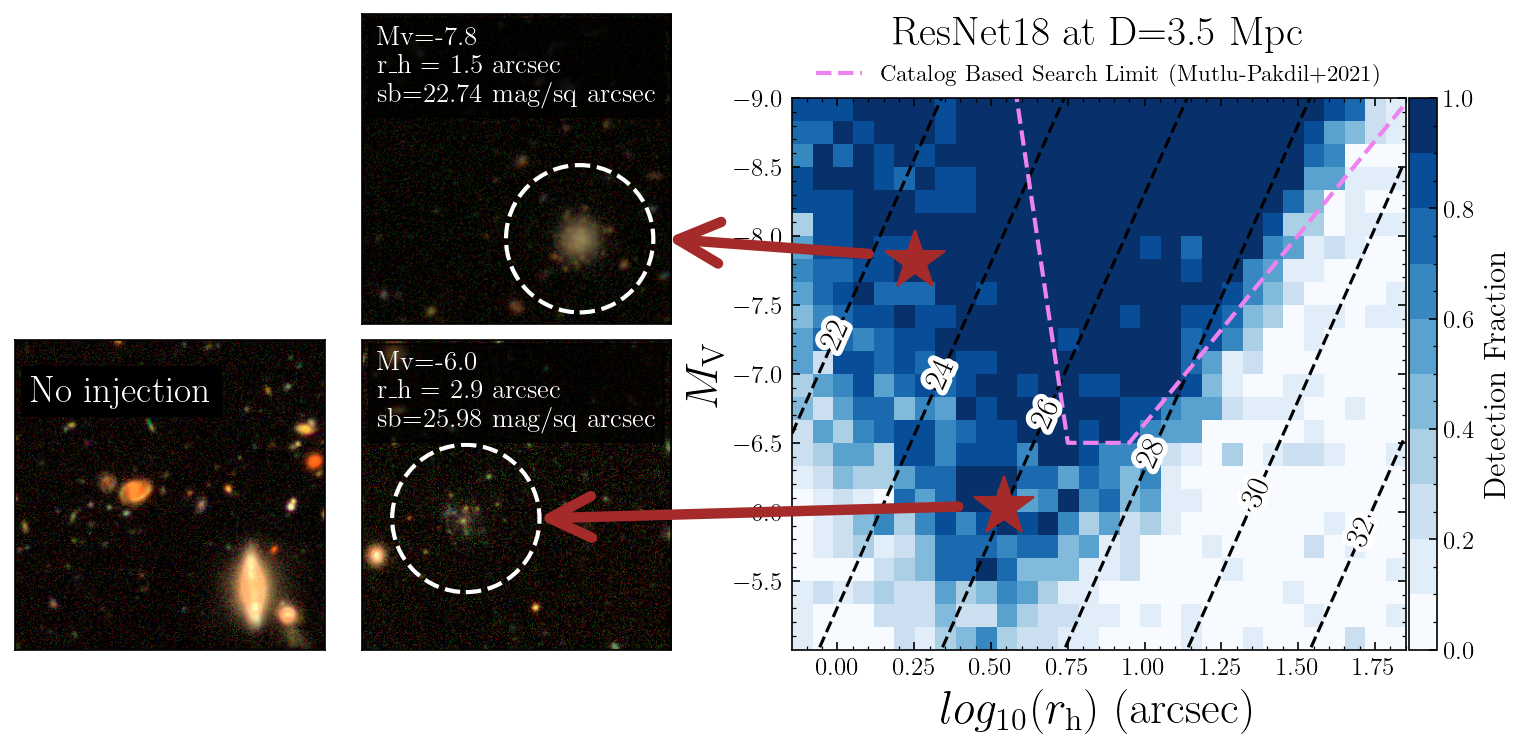}
    \caption{
    \textbf{Left:} Example $gri$-band coadd stamp (50'' wide) of HSC SSP data processed using the Rubin Science Pipelines with no source injection. 
    \textbf{Middle:} Same as left image but semi-resolved galaxies at $D_\odot \simeq 3.5$ Mpc have been injected (circled in white). 
    \textbf{Right:} Results showing detection sensitivity as a function of galaxy size ($r_h$) and brightness ($M_V$) using a supervised learning workflow with Hyrax and the Rubin Science Pipelines source injection to train a ResNet18 CNN to detect semi-resolved galaxies in HSC SSP data. The pink dashed line shows the forecasted lower limit of resolved dwarf galaxy searches at this distance from \citet[]{Mutlu-Pakdil:2021}. 
    The black dashed lines show lines of constant-surface brightness in mag per sq. arc-second.}
    \label{fig:semi-resolved}
\end{figure*}
Low-mass dwarf galaxies ($M_\star \leq 10^{7} M_\odot$) in the Local Volume ($D_\odot < 12$ Mpc) provide some of our strongest tests of galaxy formation at small scales \citep[][]{Bullock:2017,Sales:2022}, including constraints on the low-mass galaxy-halo connection \citep[e.g.][]{Nadler:2020} and their quenching either through environment \citep{Font:2022} or via feedback and reionization \citep[e.g.][]{RodriguezWimberly:2019, Rey:2020}. 
However, these objects can be difficult to identify due to their low surface brightness ($> 24$ mag arcsec$^{-2}$) and blending effects in ground-based observations \citep[][]{Mutlu-Pakdil:2021}. 
Dedicated efforts targeting local hosts have started to complete this census for satellite populations \citep[e.g.,][ and references therein]{Carlsten:2022, Drlica-Wagner:2021}. 
This task is more difficult for the field population, which is crucial for disentangling the effects of environment, requiring ML-assisted image-level searches \citep{Fielder:2025,Carlsten:2022,Li:2025}. 

These image-level searches using convolutional neural networks (CNNs) and other deep learning approaches will remain effective search methods for this next generation of surveys \citep[e.g.][]{Qu:2025}. 
Here, we present a simple implementation of using Hyrax to train a CNN to identify these rare semi-resolved dwarf galaxies.
We start with coadds from the HSC SSP survey processed with a modified version of the LSST Science Pipelines \citep[][]{10.71929/rubin/2570545}. 
From these $gri$-band coadds (as shown in the left column of Figure~\ref{fig:semi-resolved}), we use the Hyrax dataloader class to wrap the Rubin \texttt{source\_injection}\footnote{\url{https://pipelines.lsst.io/modules/lsst.source.injection/index.html}} package and easily generate realistic semi-resolved dwarf galaxies in pixel-level data for a subset of our images.
Subsequently, we split the images into smaller $50''$ postage stamps that form our labeled training set. 
The middle column of Figure~\ref{fig:semi-resolved} shows two of these objects injected into HSC data. 

Then, we use \texttt{hyrax.train()} to conduct supervised training of a ResNet18 CNN. 
The results of this training evaluated on a separate validation dataset are shown in the right column of Figure~\ref{fig:semi-resolved}, where we plot detection fraction as a function of the size ($r_h$) vs.\ luminosity ($M_V$) for dwarf galaxies at a distance of $D_\odot \simeq 3.5$ Mpc. 
The pink dashed line shows the limit of ground-based resolved stellar searches \citep[][]{Mutlu-Pakdil:2021}, where objects below and to the left of this line would be undetected. 
We also show black dashed lines of constant surface brightness. 
The red stars show the locations of the two semi-resolved objects highlighted in the middle column. 
Additionally, our false-positive rate of $0.2\%$ means that visual inspection of the candidates is tractable at the LSST level. 
This model demonstrates effective detection of semi-resolved galaxies in HSC data processed with the Rubin Science Pipelines, suggesting it is well-positioned for extension to future LSST observations.  
In the near future, we plan to implement the pre-processing steps of other semi-resolved searches \citep{Fielder:2025, Carlsten:2022}
in the LSST environment, and possibly develop differentiable diffusion models for our image-level injections. 

\section{Summary \& Future Direction} \label{sec:summary}
Despite the rapid growth of ML in astronomy over the past decade, the infrastructure supporting these efforts remains largely fragmented, with individual astronomers and groups developing bespoke pipelines that are often difficult to maintain, extend, or reuse. In this paper, we present Hyrax, an open-source, modular, and extensible framework that provides astronomy-aware infrastructure for the full ML lifecycle: from data acquisition and model training to inference and experiment comparison. 

Hyrax also addresses a critical infrastructure gap that is becoming more pressing as foundation models and unsupervised representation learning grow increasingly prevalent in astronomy: the lack of general-purpose, survey-agnostic tools for exploring and interrogating learned latent spaces. To this end, Hyrax provides integrated vector databases for scalable similarity search and interactive two- and three-dimensional visualization tools that link latent-space structure back to the underlying imaging and catalog data---enabling the kind of rapid, iterative inspection that is essential for extracting scientific insight from high-dimensional embeddings. 

Hyrax is intentionally agnostic to the specific scientific application or learning paradigm; by layering domain-specific functionality on top of established open-source tools, it aims to let astronomers spend their time on discovery rather than on software engineering. We demonstrated this versatility through five science applications in \S\ref{sec:applications}: unsupervised representation learning on $\sim 4 \times 10^5$ Rubin DP1 ECDFS galaxies, where iterative latent-space exploration and similarity search surfaced new galaxy merger and low-surface-brightness candidates absent from cross-matched Euclid Q1 and DES LSB reference catalogs, without any labeled training data (\S\ref{sec:rubin_dp1_lsst}); a hybrid workflow combining UMAP embedding of galaxy positions and photometric redshifts with DBSCAN clustering to identify candidate gravitationally lensed arcs around galaxy clusters (\S\ref{sec:cluster_lenses}); a multimodal classification pipeline built around \texttt{AppleCiDEr}, which jointly leverages light curves, spectra, image cutouts, and metadata for early-time transient classification (\S\ref{sec:applecider}); deployment of Hyrax's supervised training and inference infrastructure within the \texttt{KBMOD} shift-and-stack pipeline for filtering false positives in distant solar system object searches (\S\ref{sec:kbmod}); and a supervised CNN trained within Hyrax to detect semi-resolved dwarf galaxies in HSC and LSST-like imaging using synthetic source injection via the Rubin Science Pipelines, achieving a false-positive rate of $0.2\%$ (\S\ref{sec:chrysomallos}).

Beyond individual science results, the unsupervised discovery application (\S\ref{sec:rubin_dp1_lsst}) yielded two methodological insights with broader implications for representation learning in new survey data. First, unsupervised models can serve as effective quality-control tools without the need for labeled training data: artifact-affected sources naturally segregated into low-density latent-space regions, and successive rounds of removal and retraining progressively drove the sample toward a cleaner set of scientifically interesting sources (Figure~\ref{fig:artifacts_island}). Second, we found that quantitative training metrics may be insufficient for model selection in discovery-oriented workflows: the validation loss did not track monotonically with the scientific coherence of the resulting representations (Figure~\ref{fig:loss_clustering}), underscoring the need for interactive tools to inspect and compare learned representations---a capability central to Hyrax's design.

With the core infrastructure established and demonstrated across diverse astronomical use cases, our development priorities going forward center on broadening community access and deepening integration with the survey ecosystem. To this end, we are working to lower the barriers for community participation through: i) integration with alert broker systems so that models can consume live alert streams; ii) support for deploying popular community-built foundation models within the framework; and iii) templates and tooling that make it straightforward for external groups to contribute models, dataset classes, and analysis components to Hyrax. Our goal is for Hyrax to serve not just as a tool for individual research groups, but as shared infrastructure for the community: a common platform on which astronomers can build, reuse, and extend each other's ML workflows as the most information-rich surveys in the history of astronomy come online.

\begin{acknowledgments}
A.G. is supported by an LSST-DA Catalyst Fellowship; this publication was thus made possible through the support of Grant 62192 from the John Templeton Foundation to LSST-DA. The opinions expressed in this publication are those of the author(s) and do not necessarily reflect the views of LSST-DA or the John Templeton Foundation.

LINCC Frameworks is supported by Schmidt Sciences. 

A.G. acknowledges support from the University of Washington eScience Institute for the UW Data Science Postdoctoral Fellowship.

The authors acknowledge support from the DiRAC Institute in the Department of Astronomy at the University of Washington. The DiRAC Institute is supported through generous gifts from the Charles and Lisa Simonyi Fund for Arts and Sciences, and the Washington Research Foundation.

This work used the Delta-GPU cluster at NCSA through allocation PHY250100 from the Advanced Cyberinfrastructure Coordination Ecosystem: Services \& Support (ACCESS) program, which is supported by U.S. National Science Foundation grants \#2138259, \#2138286, \#2138307, \#2137603, and \#2138296.

A.S., F.F.N., A.J. and M.W.C. acknowledge support from the National Science Foundation with grant numbers PHY-2117997, PHY-2308862 and PHY-2409481.

C.O.C. and A.J.C. gratefully acknowledge support from the NSF (grant No. AST-2107800).  C.O.C. gratefully acknowledges support from the NASA CSSFP (grant No. 80NSSC26K0380). 
A.J.C. gratefully acknowledges support from the Department of Energy under award DE-SC0011665.

This material is based upon work supported in part by the National Science Foundation through Cooperative Agreements AST-1258333 and AST-2241526 and Cooperative Support Agreements AST-1202910 and 2211468 managed by the Association of Universities for Research in Astronomy (AURA), and the Department of Energy under Contract No. DE-AC02-76SF00515 with the SLAC National Accelerator Laboratory managed by Stanford University. Additional Rubin Observatory funding comes from private donations, grants to universities, and in-kind support from LSST-DA Institutional Members.

This research uses services or data provided by the Rubin Science Platform at NSF-DOE Vera C. Rubin Observatory, which is jointly funded by the U.S. National Science Foundation and the U.S. Department of Energy, Office of Science.

This project used data obtained with the Dark Energy Camera (DECam), which was constructed by the Dark Energy Survey (DES) collaboration. Funding for the DES Projects has been provided by the US Department of Energy, the U.S. National Science Foundation, the Ministry of Science and Education of Spain, the Science and Technology Facilities Council of the United Kingdom, the Higher Education Funding Council for England, the National Center for Supercomputing Applications at the University of Illinois at Urbana-Champaign, the Kavli Institute for Cosmological Physics at the University of Chicago, Center for Cosmology and Astro-Particle Physics at the Ohio State University, the Mitchell Institute for Fundamental Physics and Astronomy at Texas A\&M University, Financiadora de Estudos e Projetos, Fundação Carlos Chagas Filho de Amparo à Pesquisa do Estado do Rio de Janeiro, Conselho Nacional de Desenvolvimento Científico e Tecnológico and the Ministério da Ciência, Tecnologia e Inovação, the Deutsche Forschungsgemeinschaft and the Collaborating Institutions in the Dark Energy Survey.

The Collaborating Institutions are Argonne National Laboratory, the University of California at Santa Cruz, the University of Cambridge, Centro de Investigaciones Enérgeticas, Medioambientales y Tecnológicas–Madrid, the University of Chicago, University College London, the DES-Brazil Consortium, the University of Edinburgh, the Eidgenössische Technische Hochschule (ETH) Zürich, Fermi National Accelerator Laboratory, the University of Illinois at Urbana-Champaign, the Institut de Ciències de l’Espai (IEEC/CSIC), the Institut de Física d’Altes Energies, Lawrence Berkeley National Laboratory, the Ludwig-Maximilians Universität München and the associated Excellence Cluster Universe, the University of Michigan, NSF NOIRLab, the University of Nottingham, the Ohio State University, the OzDES Membership Consortium, the University of Pennsylvania, the University of Portsmouth, SLAC National Accelerator Laboratory, Stanford University, the University of Sussex, and Texas A\&M University.

Based on observations at NSF Cerro Tololo Inter-American Observatory, NSF NOIRLab (NOIRLab Prop. ID 2019A-0337; PI: D. Trilling), which is managed by the Association of Universities for Research in Astronomy (AURA) under a cooperative agreement with the U.S. National Science Foundation.

The Hyper Suprime-Cam (HSC) collaboration includes the astronomical communities of Japan and Taiwan, and Princeton University. The HSC instrumentation and software were developed by the National Astronomical Observatory of Japan (NAOJ), the Kavli Institute for the Physics and Mathematics of the Universe (Kavli IPMU), the University of Tokyo, the High Energy Accelerator Research Organization (KEK), the Academia Sinica Institute for Astronomy and Astrophysics in Taiwan (ASIAA), and Princeton University. Funding was contributed by the FIRST program from Japanese Cabinet Office, the Ministry of Education, Culture, Sports, Science and Technology (MEXT), the Japan Society for the Promotion of Science (JSPS), Japan Science and Technology Agency (JST), the Toray Science Foundation, NAOJ, Kavli IPMU, KEK, ASIAA, and Princeton University. 

This paper makes use of software developed for the Large Synoptic Survey Telescope. We thank the LSST Project for making their code available as free software at \url{http://dm.lsst.org}

The Pan-STARRS1 Surveys (PS1) have been made possible through contributions of the Institute for Astronomy, the University of Hawaii, the Pan-STARRS Project Office, the Max-Planck Society and its participating institutes, the Max Planck Institute for Astronomy, Heidelberg and the Max Planck Institute for Extraterrestrial Physics, Garching, The Johns Hopkins University, Durham University, the University of Edinburgh, Queen’s University Belfast, the Harvard-Smithsonian Center for Astrophysics, the Las Cumbres Observatory Global Telescope Network Incorporated, the National Central University of Taiwan, the Space Telescope Science Institute, the National Aeronautics and Space Administration under Grant No. NNX08AR22G issued through the Planetary Science Division of the NASA Science Mission Directorate, the National Science Foundation under Grant No. AST-1238877, the University of Maryland, and Eotvos Lorand University (ELTE) and the Los Alamos National Laboratory.

Based, in part, on data collected at the Subaru Telescope and retrieved from the HSC data archive system, which is operated by Subaru Telescope and Astronomy Data Center at National Astronomical Observatory of Japan.

Supported by the National Science Foundation under Grants No. AST-1440341 and AST-2034437 and a collaboration including current partners Caltech, IPAC, the Oskar Klein Center at Stockholm University, the University of Maryland, University of California, Berkeley, the University of Wisconsin at Milwaukee, University of Warwick, Ruhr University, Cornell University, Northwestern University, and Drexel University. Operations are conducted by COO, IPAC, and UW.

This work has made use of the Euclid Q1 data from the {\it Euclid} mission of the European Space Agency (ESA), 2025, \url{https://doi.org/10.57780/esa-2853f3b}.

This research has made use of the Astrophysics Data System, funded by NASA under Cooperative Agreement 80NSSC21M0056.

Large language models from Anthropic (Claude) and OpenAI (ChatGPT) were used to assist in editing human-written text in this manuscript. AI-assisted coding tools---Claude Code, OpenAI Codex, and GitHub Copilot---were additionally used in the development and maintenance of the Hyrax codebase; all AI-generated contributions were reviewed and verified by human developers before merging. The specific contributions of these tools are documented in Hyrax's publicly available git history at \url{https://github.com/lincc-frameworks/hyrax}, where individual pull requests record the nature and extent of each AI-assisted change. The form of this documentation varies by tool: in some cases the complete agent conversation is preserved (e.g., \href{https://github.com/lincc-frameworks/hyrax/pull/837}{PR~\#837}); in others, a direct link to the interactive coding session is included (e.g., \href{https://github.com/lincc-frameworks/hyrax/pull/901}{PR~\#901}); and in others, the AI-generated changes are directly visible in the pull request diff (e.g., \href{https://github.com/lincc-frameworks/hyrax/pull/872}{PR~\#872}).

\end{acknowledgments}

\begin{contribution}

A. Ghosh led the overall writing and editing of this manuscript and is the corresponding author.

A. Ghosh, D. Oldag, and M. Tauraso have led the Hyrax project since its inception and are co-first-authors. D. Jones, S. Venkatesh, and G. Wang have contributed significantly to the Hyrax code base.

A. Ghosh led the writing of \S\,\ref{sec:rubin_dp1_lsst}. T. Chatchadanoraset and D. Miura also contributed to this section. 

G. Khullar led the writing of \S\,\ref{sec:cluster_lenses}. D. Berry also contributed to this section.

A. Sasli led the writing of \S\,\ref{sec:applecider}. M. Coughlin and F. Fontinele Nunes also contributed to this section.

A. J. Connolly \& M. West led the writing of \S\,\ref{sec:kbmod}. H. M. Tohfa also contributed to this section.

P. S. Ferguson led the writing of \S\,\ref{sec:chrysomallos}. 

W. Beebe, D. Branton, S. Campos, N. Caplar, C. O. Chandler, M. Dai, M. DeLucchi, A. Junell, J. Kubica, K. Malanchev, R. Mandelbaum, S. McGuire, N. Morato, I. Pasha, D. Taranu, and T. Zhang contributed to the overall design and development of the Hyrax codebase and the projects presented herein, and provided feedback on this manuscript.


\end{contribution}

%
\facilities{Rubin:Simonyi, Rubin:USDAC, Euclid, Blanco, PO:1.2m, Subaru}


\software{
    Numpy \citep{Harris2020ArrayNumPy},
    Scipy \citep{scipy},
    Astropy \citep{2013A&A...558A..33A, 2018AJ....156..123A, astropy:2022},
    Pandas \citep{pandas},
    Matplotlib \citep{Hunter2007Matplotlib:Environment},
    LINCC Frameworks Python Project Template \citep{oldag2024python},
    PyTorch \citep{Ansel2024PyTorchCompilation},
    TorchVision (\url{https://github.com/pytorch/vision}),
    PyTorch Ignite (\url{https://github.com/pytorch/ignite}),
    MLflow \citep{mlflow},
    Optuna \citep{Akiba2019Optuna},
    umap-learn \citep{McInnes2020UMAP:Reduction},
    HuggingFace Datasets \citep{lhoest2021datasets},
    schwimmbad \citep{schwimmbad2017},
    pooch \citep{uieda2020pooch},
    tqdm \citep{dacostaluis2019tqdm},
    HoloViews \citep{stevens2015holoviews},
    Cython \citep{behnel2011cython},
    LanceDB/Lance \citep{pace2025lance},
    TensorBoard (\url{https://github.com/tensorflow/tensorboard}),
    ChromaDB (\url{https://github.com/chroma-core/chroma}),
    Qdrant (\url{https://github.com/qdrant/qdrant}),
    ONNX (\url{https://github.com/onnx/onnx}),
    ONNX Runtime (\url{https://github.com/microsoft/onnxruntime}),
    ONNXScript (\url{https://github.com/microsoft/onnxscript}),
    Plotly (\url{https://plotly.com}),
    PyArrow/Apache Arrow (\url{https://arrow.apache.org}),
    Bokeh (\url{https://bokeh.org}),
    Datashader (\url{https://datashader.org}),
    nested-pandas (\url{https://github.com/lincc-frameworks/nested-pandas}),
    Pydantic (\url{https://docs.pydantic.dev}),
    psutil (\url{https://github.com/giampaolo/psutil}),
    colorama (\url{https://github.com/tartley/colorama}),
    more-itertools (\url{https://github.com/more-itertools/more-itertools})
}

\appendix

\section{Code, Documentation \& Tutorials} \label{sec:ap:code}
Hyrax has been developed as an open-source framework since its inception. Below, we outline links to Hyrax's source code, documentation, and tutorials.

\begin{itemize}
    \item Source Code: \url{https://github.com/lincc-frameworks/hyrax}
    \item Documentation: \url{https://hyrax.readthedocs.io/}
    \item Tutorials: \url{https://hyrax.readthedocs.io/en/stable/science_examples.html}
\end{itemize}

Hyrax maintains a unit test suite covering its core modules, with continuous integration run automatically on every pull request across multiple Python versions and code coverage reported via Codecov. The framework is built on the LINCC Frameworks Python Project Template \citep{oldag2024python}, which enforces consistent code style via \texttt{ruff} linting and pre-commit hooks, and uses semantic versioning to maintain a stable public API.

Since Hyrax is a living code repository, which we expect to keep changing with time, we have created a ``frozen" version of the code at the time of writing this article. This version is tagged as release \texttt{v0.8.1} in the above-mentioned GitHub repository and on \href{https://pypi.org/project/hyrax/#history}{PyPI}.

\bibliography{references-2}
\bibliographystyle{aasjournalv7}

\end{document}